\documentclass[aps,onecolumn,prl,floatfix]{revtex4}
\usepackage{amsmath} \usepackage{amsfonts} \usepackage{amssymb}
\usepackage{bbm}

\usepackage{graphics}

\usepackage{graphicx}

\usepackage[font=normal]{subfig}
\usepackage{txfonts}
\usepackage{color}

\usepackage{ulem}

\begin{document}

\title{Ferroelectric $\pi$-stacks of molecules with the energy gaps in the sunlight range}

\author{Pawe\l{} Masiak and Ma\l{}gorzata~Wierzbowska}

\affiliation{
Institute of Physics, Polish Academy of Sciences\\
Aleja Lotnikow 32/46, PL-02668 Warsaw, Poland}


\begin{abstract}
Ferroelectric $\pi$-stacked molecular wires for solar cell applications 
are theoretically designed, in such a way that their energy gaps fall within 
visible and infrared range of the Sun radiation. 
Band engineering is tailored by a modification
of the number of the aromatic rings and via a choice of the number and kind of the dipole groups.
The electronic structures of molecular 
wires and the chemical character of the electron-hole pair 
are analyzed within the density functional theory (DFT) framework
and the hybrid DFT approach by means of the B3LYP scheme.
Moreover, it is found that one of the advantageous properties of these systems \--
namely the separate-path electron and hole transport \-- reported earlier, 
still holds for the larger molecules, due to the dipole selection
rules for the electron-hole generation, which do not allow the lowest optical transitions
between the states localized at the same part of the molecule.
\end{abstract}

\maketitle

\section{Introduction}

The ability to absorb a wide range of Sun spectrum and convert this energy into 
the voltage between the electrodes is a key factor of the efficient solar battery. 
Therefore, the optically active materials ideally should be the composite materials 
of small-, middle- and wide-bandgap semiconductors, in order to cover the whole radiation range
from the soft ultaviolet (350 nm) to the far infrared of a sunset (1400 nm).
This possibility is offered by the multilayers of planar molecules or arrays of molecular wires. 

Recently, the so-called covalent organic frameworks (COF) attract an attention due to their
special optical, transport, and catalytic properties, 
as well as easy fabrication \cite{stang1,stang2,bein1,bein2}.
From a point of view of the photovoltaics, stacks of the COFs are similar
to the bulk or integrated heterojunctions \cite{bein2,bulk}.   
By a change of the planar bonds between 
the molecules from the covalent to hydrogen bonds,
one can restrict the electronic transport to the direction across the layers,
i.e. along the stacks, while the electronic transport
within the planes shall be suppressed.
This is advantageous for the solar batteries, 
where the planar current would cause only a dissipation of an energy.
Therefore, in the previous works \cite{go1,go2}, we studied the layers of molecules 
with the COOH terminal groups as the connecting parts for building the networks.
The COOH group possesses also a small dipole moment.

The ferroelectrically ordered molecules, composed of benzene rings and two dipole groups 
(COOH and CH$_2$CN), arranged in the $\pi$-type stacks of layers or the molecular wires,   
show many appealing effects \cite{go1,go2}. 
In particular, the energy levels of the subsequent layers (or molecules in a stack) 
are aligned in a cascade, and this holds for the valence and conduction band as well \cite{go1,go2,as}. 
Each layer is simultaneously a donor and acceptor of electrons and holes, 
depending on a direction of the carrier motion \cite{go1}. 
The excitons in such layers are localized and have the charge-transfer character   
from the dipole group to the aromatic central ring. The electric field generated by the  
ferroelectrically ordered dipole groups leads to a polarization 
which is induced at electrodes.
This effect for the graphene sheets, chosen as the electrodes, causes a change of the work function by 
$\pm$1.5 eV for the anode and cathode, respectively \cite{go1}. Moreover, the electrons and   
holes move across such $\pi$-stacks along different paths: the electrons through the central rings
and holes between the dipole groups \cite{go2}. The carrier mobilities, obtained with the   
relaxation time estimated due to the elastic scattering and ionic intrusions, are higher than   
those in the organometal halide perovskites \cite{go2,SS,snaith}. 
For all the above reasons, it is interesting to investigate further properties of 
the ferroelectric molecular layers and stacks, in order to bring these systems closer to 
the experimental and industrial interest \cite{advmat}.

\begin{figure*}
\begin{tabular}{cccc}
 b1(COOH)$_3$  &  b1(CH$_2$CN)$_3$  &  b1(CH$_2$CF$_3$)$_3$  &  b1(COOH,CH$_2$CN)$_3$ \\

\includegraphics[scale=0.15,angle=0.0]{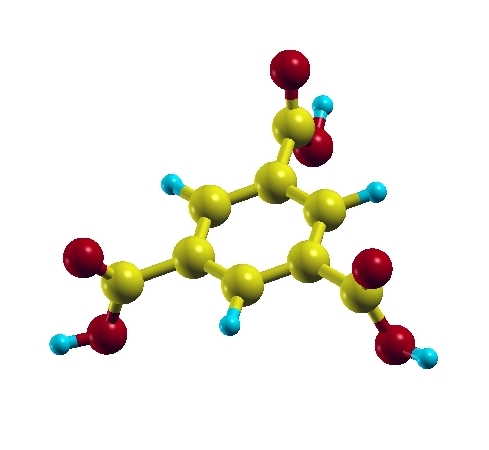}\label{} &
\includegraphics[scale=0.15,angle=0.0]{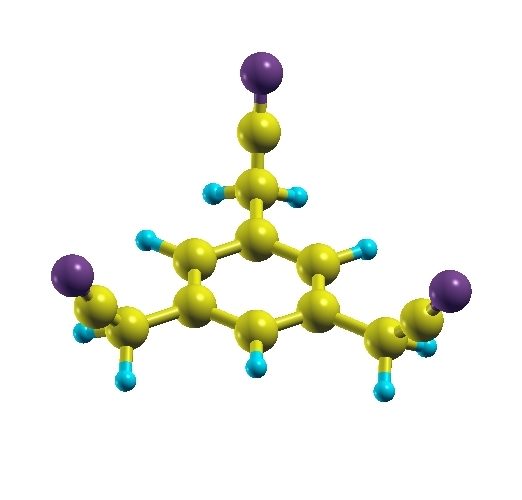}\label{} &
\includegraphics[scale=0.15,angle=0.0]{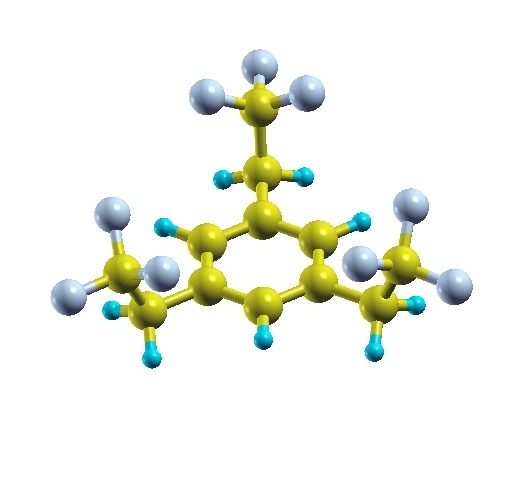}\label{} &
\includegraphics[scale=0.15,angle=0.0]{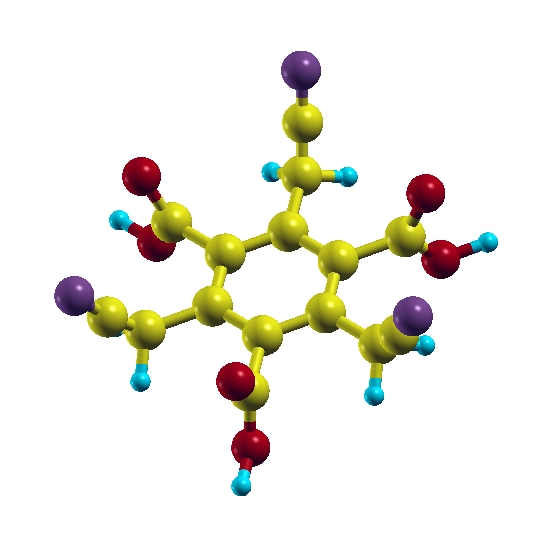}\label{} \\[0.3cm]

\multicolumn{2}{c}{b5(CH$_2$CN)$_{10}$}  &  \multicolumn{2}{c}{b9(COOH)$_4$}  \\

\multicolumn{2}{c}{\includegraphics[scale=0.2,angle=0.0]{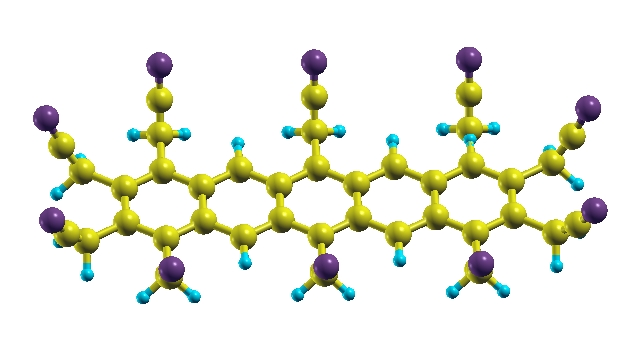}\label{}} &
\multicolumn{2}{c}{\includegraphics[scale=0.17,angle=0.0]{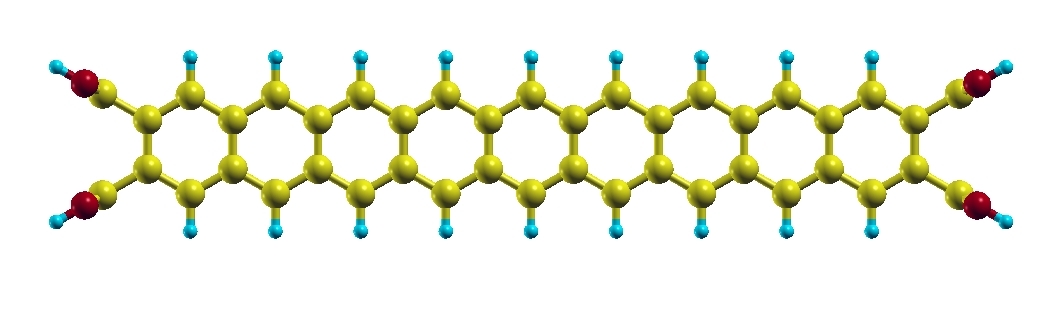}\label{}} \\[0.3cm]

\multicolumn{4}{c}{b17(COOH)$_{8}$} \\

\multicolumn{4}{c}{\includegraphics[scale=0.3,angle=-0.5]{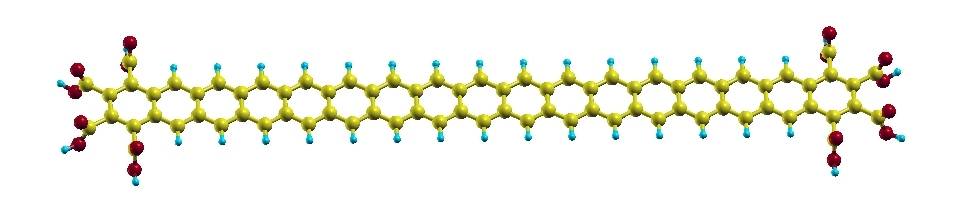}\label{}}

\end{tabular}
\caption{Atomic structures of some of the studied molecules and their short names.}
\end{figure*}

In this theoretical work, we focus on the rules which govern the bandgap change. 
The energy gap should fall into an interesting for us range of the Sun radiation.    
Recently, a similar system (to the cases investigated by us) has been theoretically and 
experimentally investigated, namely 2D imine polymer \cite{lublin}.
The authors revealed that a bandgap tunning by expanding a conjugation of the backbone 
of the aromatic diamines is possible in this material \cite{lublin}.
It is also well known that the bandgap decreases with 
a size of a system \cite{taiwan,usa-china}. However, without calculations, 
the exact value of the energy gap is difficult to predict, as well as
its dependence on a symmetry and the edge termination \cite{usa-china,belgium}. 

We have begun our study by choosing a type of molecules, which could be most promising for our purpose. 
Fig. 1 presents some of the molecules which are analyzed in this work. 
The collection of the geometries of all other studied systems, not presented in figures here, 
is included in the supporting information. We have chosen three dipole
groups: COOH, CH$_2$CN, CH$_2$CF$_3$, and their combinations in various repetitions.
The size of the mesogenic aromatic part was enlarged linearly.
The number of the benzene-type rings is given in our notation by an integer number following 
the "b" letter.

Firstly, we investigated an effect of a number of the aromatic rings and number of
dipole groups on a size of the energy gap; it means, the difference between the energetic positions 
of the lowest unoccuped and the highest occupied molecular orbital (LUMO-HOMO).
Secondly, we checked an effect of mixing various chemical groups 
as the terminal dipoles in one molecule.
We estimated also an effect of the $\pi$-stacking. 
Finally, we analyzed how the molecular modiffications 
\-- which were done in order to tailor the bandgap size \--
affect the exciton (electron-hole) character and the separation
of the charge carrier paths.  
  
\section{Theoretical methods}

The molecular calculations have been performed with the Gaussian code \cite{gauss}, 
using the correlation-consistent valence double-zeta atomic basis set with
polarization function cc-pVDZ \cite{gaubas}. 
Molecular geometries were optimized with the hybrid-functional method in the B3LYP flavor \cite{b3lyp}, 
which mixes the density functional theory (DFT) \cite{dft} in the BLYP \cite{blyp1,blyp2} parametrization
with the Hartree-Fock exact exchange in 80$\%$ and 20$\%$, respectively.
The optimized atomic structures were used to build the molecular wires. 

Further calculations for the molecules and 1D structures have been performed with the 
{\sc Quantum ESPRESSO} suite of codes \cite{qe}. This package is based
on the plane-wave basis set and the pseudopotentials for the core electrons.
The normconserving pseudopotentials were used with the energy cutoff  
for the plane-waves set to 35 Ry. Moreover, some of the results, such 
as optimized  intermolecular distances, were checked to be the same as using the
larger energy cutoff of 45 Ry. 
The intermolecular distances were obtained within the local density approximation (LDA),
since it is known to give better geometries than the generalized gradient approximation (GGA). 
The LDA-optimized separations between molecules in the stack were then used for the B3LYP calculations for the
molecular wires.
The uniform Monkhorst-Pack k-points mesh in the Brillouin zone \cite{mp} was chosen 
for 1$\times$1$\times$10 for the wires. For the B3LYP scheme, we used the meshes
1$\times$1$\times$9 and 1$\times$1$\times$3 for the k- and q-point grids, respectively. 

In order to obtain the band structures projected onto the local groups of atoms,
we employed the wannier90 package \cite{w90}, which interpolates bands using
the maximally-localized Wannier functions \cite{wan,RMP}. The same tool has been used
for the calculations of the dipole moment, 
which can be obtained from the positions of the maximally-localized Wannier
centers, $r_n$, using the formula \cite{polar} 
\begin{equation}
d = \sum_a Z_a R_a - \sum_n r_n
\end{equation}
where $Z_a$ and $R_a$ are the atomic pseudopotential charge and its position, correspondingly,
and indexes $a$ and $n$ run over the number of atoms and Wannier functions, respectively.

\section{Results}

\subsection{Dipole moment}

There are a number of important implications of using the dipole groups: i) the hydrogen
bonds via the COOH groups within the planes restrict the electronic transport and
dissipation of energy in the directions perpendicular to the photovoltaic path,
ii) the polarization generated across the solar device orders 
the energy levels in a cascade \cite{go1}, 
iii) the electronic transport along the $\pi$-stacks is restricted to the $\pi$-conjugated rings 
for the electrons, and holes move between the dipole groups \cite{go2}, 
iv) an adsorption of the optically active molecules at the surfaces of 
the transparent conductive oxides, used as the electrodes, leads to the
high power conversion when it is realized with the COOH group \cite{adsor}. 
Therefore, it is usefull to examine various dipole groups for their impact on the energy gaps.
In table 1, the dipole moments in the direction parallel to the photovoltaic transport 
for all studied groups attached to one aromatic ring are collected.     

\begin{table}[h]
\caption{Dipole moment (in Debye) of benzene with chosen polar groups \-- a component
perpendicular to the aromatic ring (D$_z$) \--  
obtained from the LDA calculations and the Wannier-functions spreads according to Eq. 1.}
\begin{tabular}{lc}
\hline \\[-0.2cm]
 Molecule &  D$_z$ \\[0.1cm]
\hline \\
 b1(COOH)$_1$                   & -0.07  \\
 b1(CH$_2$CN)$_1$               & -0.59  \\
 b1(CH$_2$CF$_3$)$_1$           & -0.35  \\
 b1(COOH,CH$_2$CN)$_3$ $\;\;\;$ & -2.06  \\[0.1cm] 
\hline
\hline
\end{tabular}
\label{dip}
\end{table}

\begin{figure*}[b]
\vspace{5mm}
\centerline{
\includegraphics[scale=0.35]{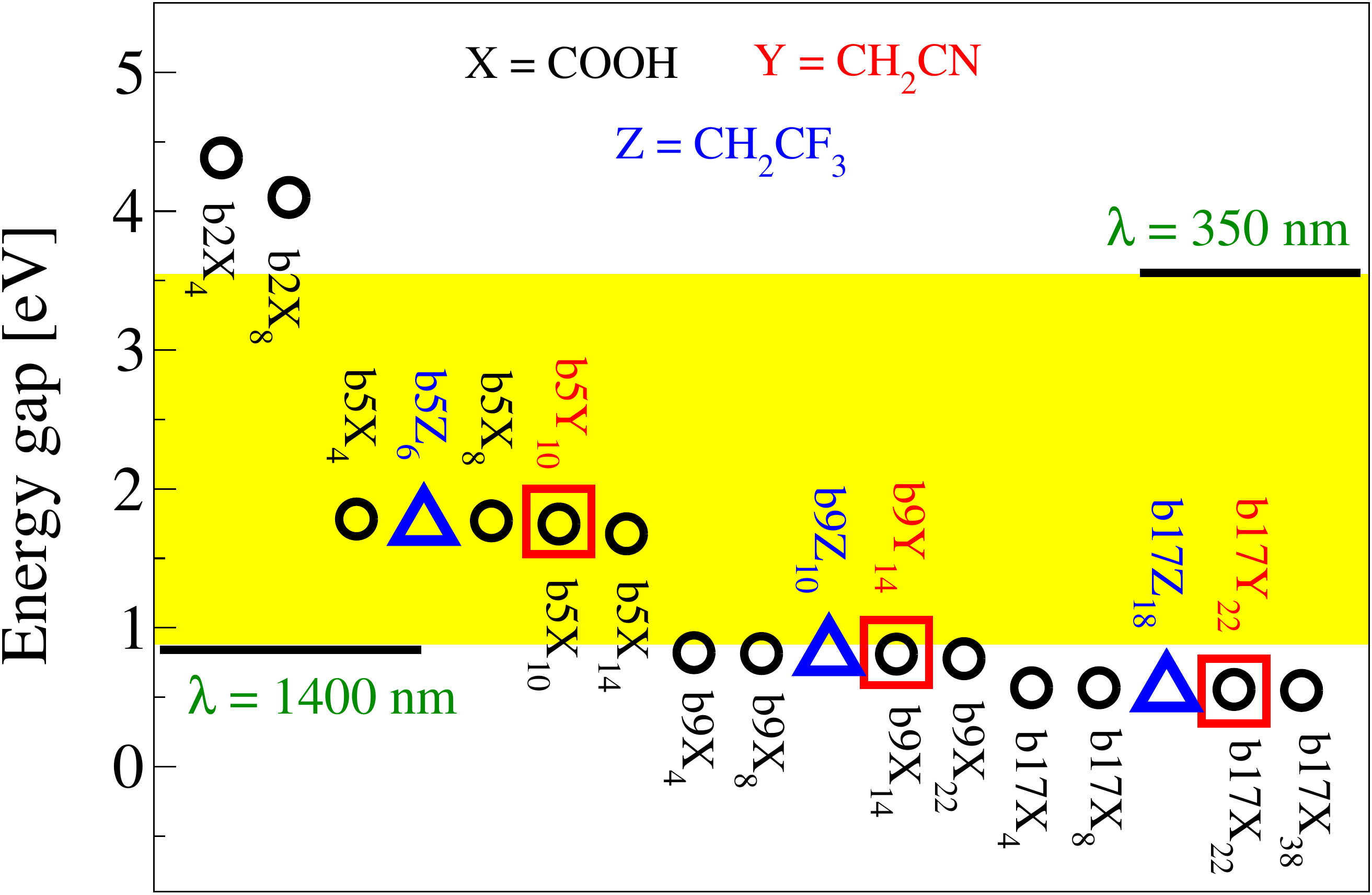} }
\vspace{5mm}
\caption{The LUMO-HOMO energies of the isolated molecules 
with variable number of benzene rings
and three dipole groups: COOH (black circles), CH$_2$CN (red squares), CH$_2$CF$_3$ (blue triangles)
\-- obtained with the B3LYP method. The solar spectrum range is marked in yellow.
All calculations have been performed with the Gaussian code.}
\label{b2}
\end{figure*}

\subsection{LUMO-HOMO energy differences}

A collection of molecules with various lengths of the aromatic chains is gathered in Fig. 2.
The energy gaps are presented on a scale of the Sun radiation activity
from the soft ultraviolet (350 nm) to the far infrared of a cloudy sky (1400 nm).  
The impact of various dipole groups on the LUMO-HOMO energy difference 
does not change with the group types, 
when they are attached to the long chains (above five benzene rings). Moreover, increasing 
a number of the dipole groups does not change the energy gap for the longer molecules.
Due to the lack of this effect, we can focus on the transport 
properties when a design of the dipole groups is considered. 

It is a well known fact, that the energy gaps obtained with the density functional theory \--
both in the local density approximation (LDA) and the generalized gradient approximation (GGA)
\-- are underestimated, while the energy gaps from a pure Hartree-Fock method are 
largely overestimated. Thus, we used the hybrid-functional scheme by means of the B3LYP
functional which contains 20$\%$ of the exact exchange.
The presented series of the energies could be 
an approximation of the optical gaps \-- when just a tendency of the size and chemical group 
effect on the gaps is studied. The experimental data are usually closer
to the combined GW+BSE approach taking into account both the quasi-particle and
the excitonic binding energy effects \cite{GWall,BSE}. 
However, from the presented data, 
it is obvious that a realization of the highly efficient \-- from the light-absorption 
point of view \-- matrices of molecular stacks is possible. Especially, if one combines
the layers with columns of various molecules, which have different size and different absorption profiles.

\subsection{Impact of the $\pi$-stacking for the energy gaps}

The absorption efficiency is correlated with the thickness of the photoactive layer
(Beer-Lambert Law),
which cannot be too small making the material to be transparent \cite{abseff}. 
Hence, all 2D structures for the solar cells should be examined for their properties
across the layers. Stacking causes the band dispersions in the direction perpendicular
to the slab. This bandwidth broadening, in turn, acts for the closure of the energy gap.
In Table 2, we present an effect of the wire formation on the band gap, and compare 
the LDA and B3LYP computational methods for chosen systems.    

\begin{table*}[h]
\caption{Energy gaps (in eV) of the isolated molecules 
and molecular wires, obtained with
the LDA and B3LYP methods. The last column displays the intermolecular distances (in $\AA$)
in the wires (Dist.) \-- obtained with the LDA scheme and used also for the B3LYP calculations.
All calculations have been performed with the QE code.}

\begin{tabular}{lccccccc}
\hline \\[-0.2cm]
 molecule & \multicolumn{2}{c}{$\;\;\;\;\;$ $E_{gap}$ (isolated) $\;\;\;\;\;\;$} &
\multicolumn{2}{c}{$\;\;\;\;$ $E_{gap}$ (wire) $\;\;\;\;\;\;$} & 
\multicolumn{2}{c}{$\;\;\;\;$ $\Delta E_{gap}$ (wire-isolated) $\;\;\;\;\;\;$} & 
 Dist. (wire) \\ 
                   & $\;\;\;\;$ LDA $\;\;$ & $\;\;$ B3LYP $\;\;\;\;$ & 
                     $\;\;\;\;$ LDA $\;\;$ & $\;\;$ B3LYP $\;\;\;\;$ &
                     $\;\;\;\;$ LDA $\;\;$ & $\;\;$ B3LYP $\;\;\;\;$ &  $\;\;\;\;$ LDA $\;\;$  \\[0.1cm]
\hline \\
 benzene                        &  5.119  &  6.717   &  4.026  &   5.452  &  -1.093  &  -1.265  & 3.8 \\
 b1(COOH)$_3$                   &  4.204  &  6.199   &  4.037  &   6.016  &  -0.167  &  -0.183  & 5.1 \\
 b1(COOH)$_6$ $\;\;\;$          & 3.956   & 5.875    & 3.857   & 5.813    & -0.099   & -0.062   & 5.1 \\
 b1(CH$_2$CN)$_3$               &  4.293  &  6.074   &  3.986  &  5.692   &  -0.307  & -0.382   & 4.4 \\
 b1(COOH,CH$_2$CN)$_3$ $\;\;\;$ &  3.611  &  5.503   &  2.963  &   5.050  &  -0.648  &  -0.452  & 4.6 \\
 b1(CH$_2$CF$_3$)$_3$           &  4.759  &  6.359   &  4.607  &   6.119  &  -0.152  &  -0.239  & 5.2 \\
 b1(COOH,CH$_2$CF$_3$)$_3$          & 4.084  & 5.958  & 3.946  & 5.871 & -0.138  &  -0.087  & 5.2 \\
 b1(COOH,CH$_2$CN,CH$_2$CF$_3$)$_1$ & 4.249  & 6.196  & 4.025  & 5.989 & -0.224  & -0.207   & 5.0  \\
 b5(COOH)$_4$           &  0.822  &  1.783   &  0.738  &   1.670  &  -0.084  &  -0.113  & 5.1 \\[0.1cm]
\hline
\hline
\end{tabular}
\label{occ}
\end{table*}

The last column in Table 2 displays the intermolecular distances obtained with the LDA method.
The separation of benzenes, by 3.8 $\AA$, is not much larger than that of the graphene multilayers, 
which is around 3.4 $\AA$ \cite{graphene}. 
The most distant are molecules with the CH$_2$CF$_3$ groups, of 5.2 $\AA$. This is due to the fact
that they are the largest of all atomic groups attached to the rings studied here, 
and the F atoms do not attract the H atoms at the bottom of the upper neighbor.
The COOH groups are the smallest here, but separations of the molecules
terminated by them are also large \-- of 5.1 $\AA$ \-- 
because the oxygen atoms from the neighboring rings repel each other. 
The CH$_2$CN groups, although they are also quite large, attract each other between the neighboring
rings. This is because N and C tend to "exchange" hydrogen, and this effect leads to the smallest
intermolecular distances, of 4.6 $\AA$.  
The separations of molecules in a wire rule an effect of the bandgap size in the stack;
this effect is the strongest for the benzene wires and molecules containing the CH$_2$CN groups.
It is interesting to note, that an addition of COOH to the rings with other dipole groups 
weakens an effect of stacking on the band gap. This holds 
even at the same intermolecular distance (see for instance b1(CH$_2$CF$_3$)$_3$ and
b1(COOH,CH$_2$CF$_3$)$_3$, or b1(COOH)$_3$ and b1(COOH)$_6$ in Table 2).
The origin of this effect will be more clear in the next subsection. 

Comparison of the LDA and B3LYP approaches for the isolated molecules and wires, usually, 
exhibits a bit stronger effect of the stacking on the band gap for the hybrid-functional scheme.
Also the origin of this effect is similar to that of an addition of COOH, and relies on the order
of the energy levels. Because the experimental results for the band gaps of the molecules studied 
here do not exist yet, it is not easy to determine how good is the B3LYP method in this case. 
However, one could expect that this method will work similar as in the case of benzene. 
In order to evaluate effectiveness of the B3LYP method used in our study, 
we compared energies for benzene calculated by different theoretical methods with experimental result. 
The energy gap of the benzene molecule (isolated) of 6.72 eV by means of the B3LYP should be compared with 
the value from the GW approach, of 10.5 eV \cite{Louie}, 
because both approaches do not take into account the excitonic effects.
However, the measured optical gap is around 3.6 eV \cite{opt}, 
due to the effect of large exciton binding energy in small molecules.
This excitonic effect can be theoretically obtained from the difference of the gaps obtained
with the GW and the GW+BSE (BSE means Bethe-Salpeter equation). 
For benzene, the GW+BSE band gap is around 3.1 eV \cite{BSE}.
Although the B3LYP scheme is much simpler then the GW and GW+BSE methods, and designed for 
the fundamental gap only, its results are closer to the optical absorption
than these of the GW aproach. Therefore, we expect that the energy gaps obtained
by us with the B3LYP method 
show the same trends in a series of similar molecules \-- which differ
only with a size or number or type of the dipole groups \-- as 
the optical measurements.

\subsection{Order of the energy levels}

The characters of the highest occupied and lowest unoccupied states determine
the excitonic radius and binding energy, the oscillator strength of the absorption of light,
as well as the electronic transport properties. 

\begin{figure*}[h]
\begin{tabular}{cccc}
\includegraphics[scale=0.17]{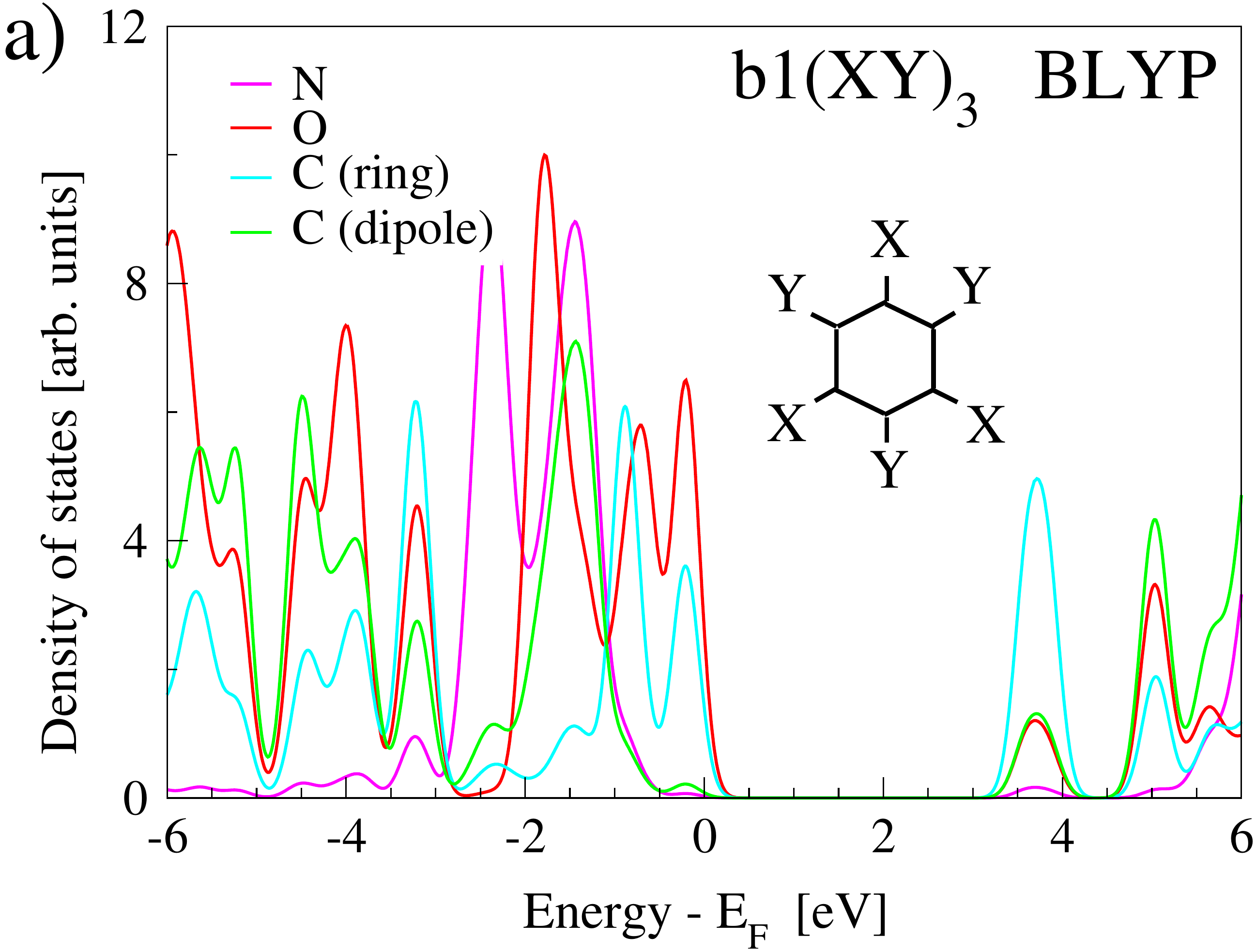} \hspace{0.2cm} &
\includegraphics[scale=0.17]{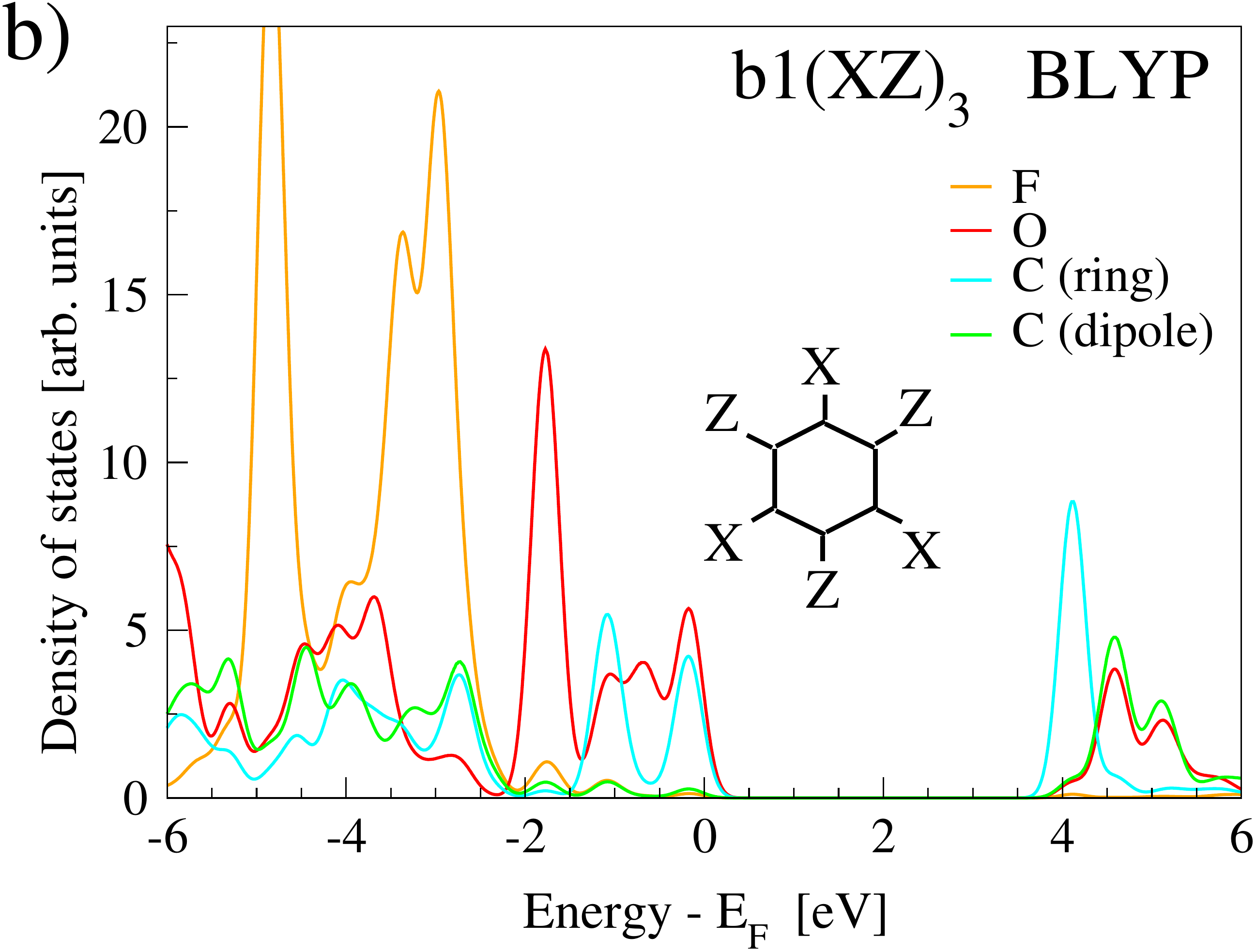} \hspace{0.2cm} &
\includegraphics[scale=0.17]{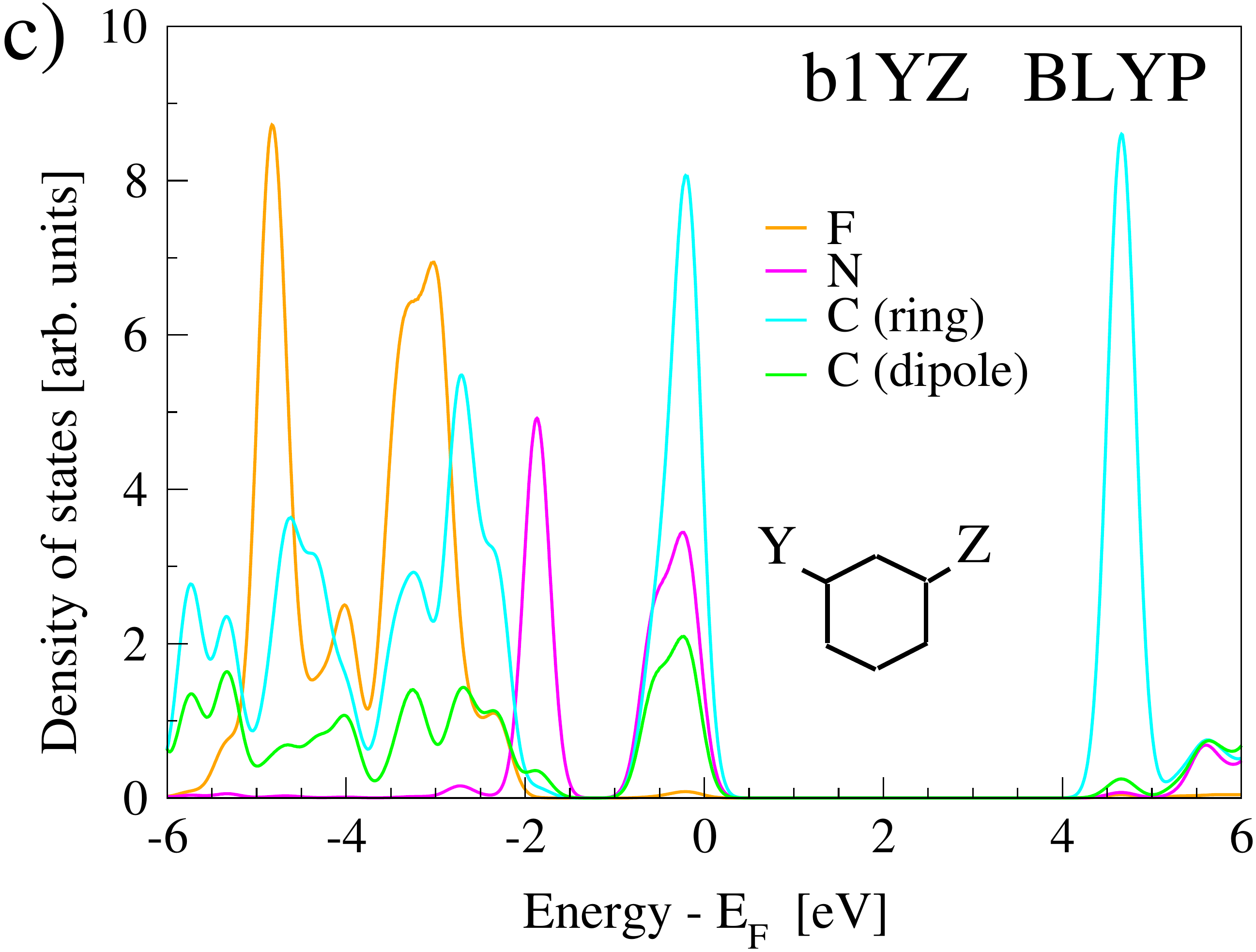} \\[0.3cm] 

\includegraphics[scale=0.17]{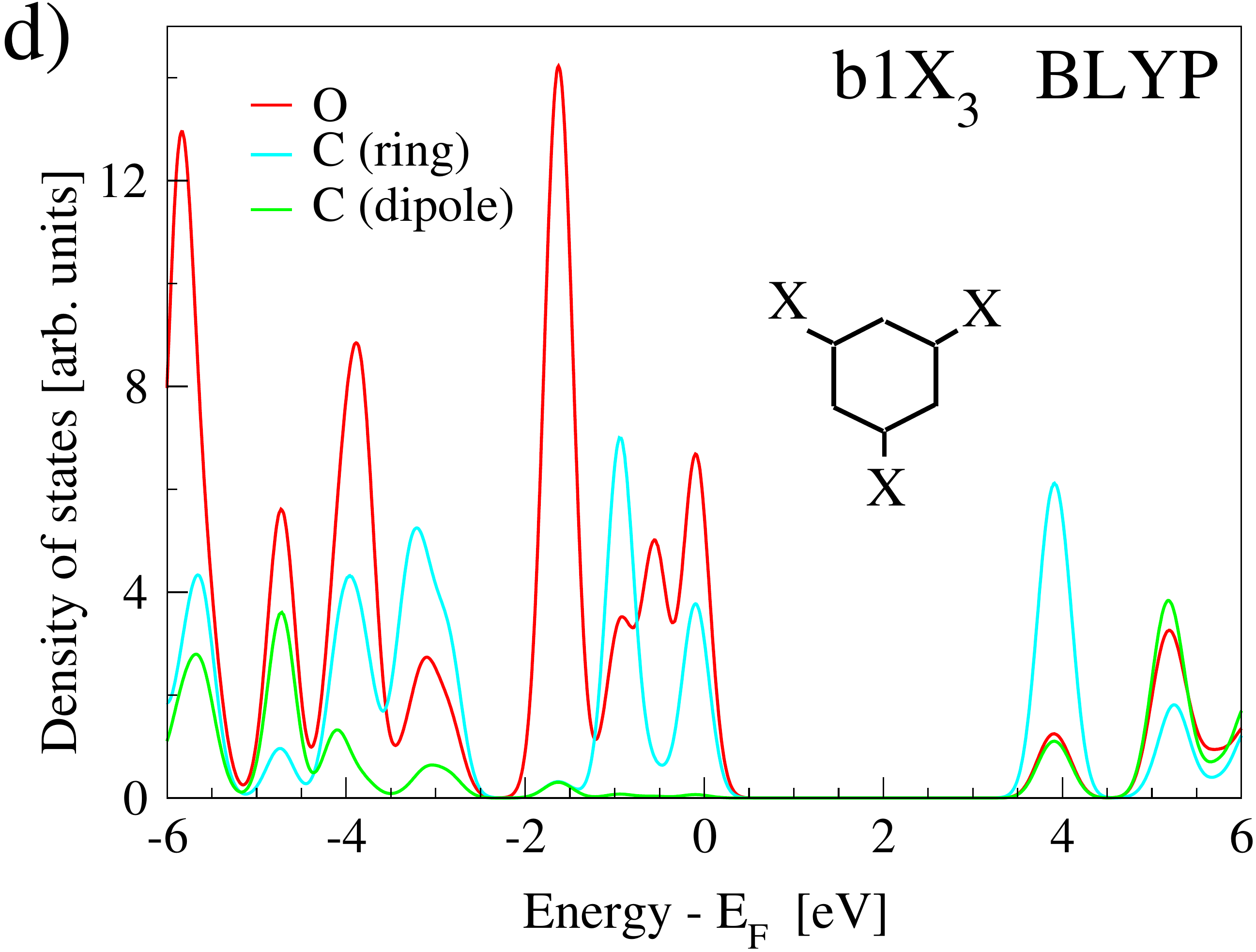} \hspace{0.2cm} &
\includegraphics[scale=0.17]{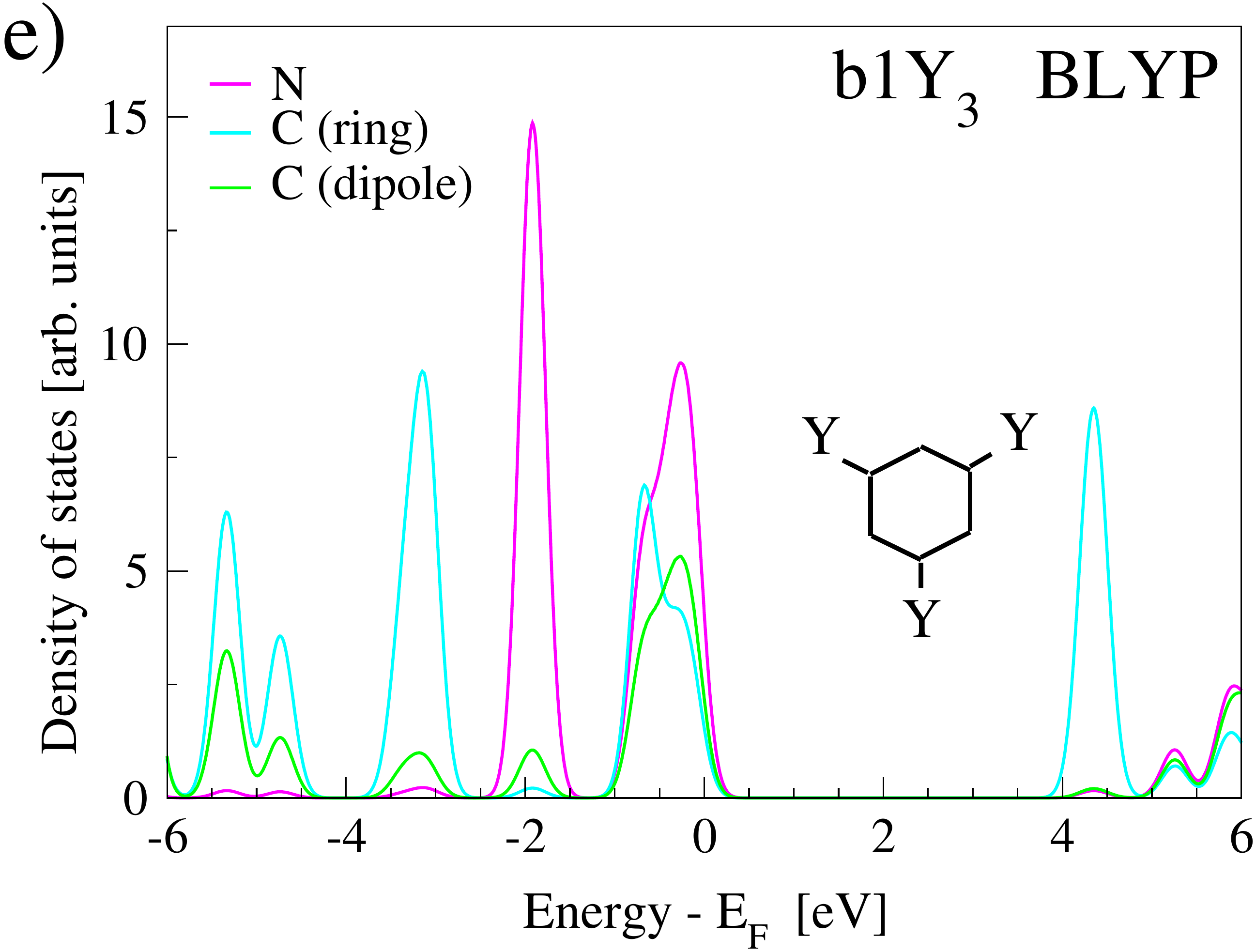} \hspace{0.2cm} &
\includegraphics[scale=0.17]{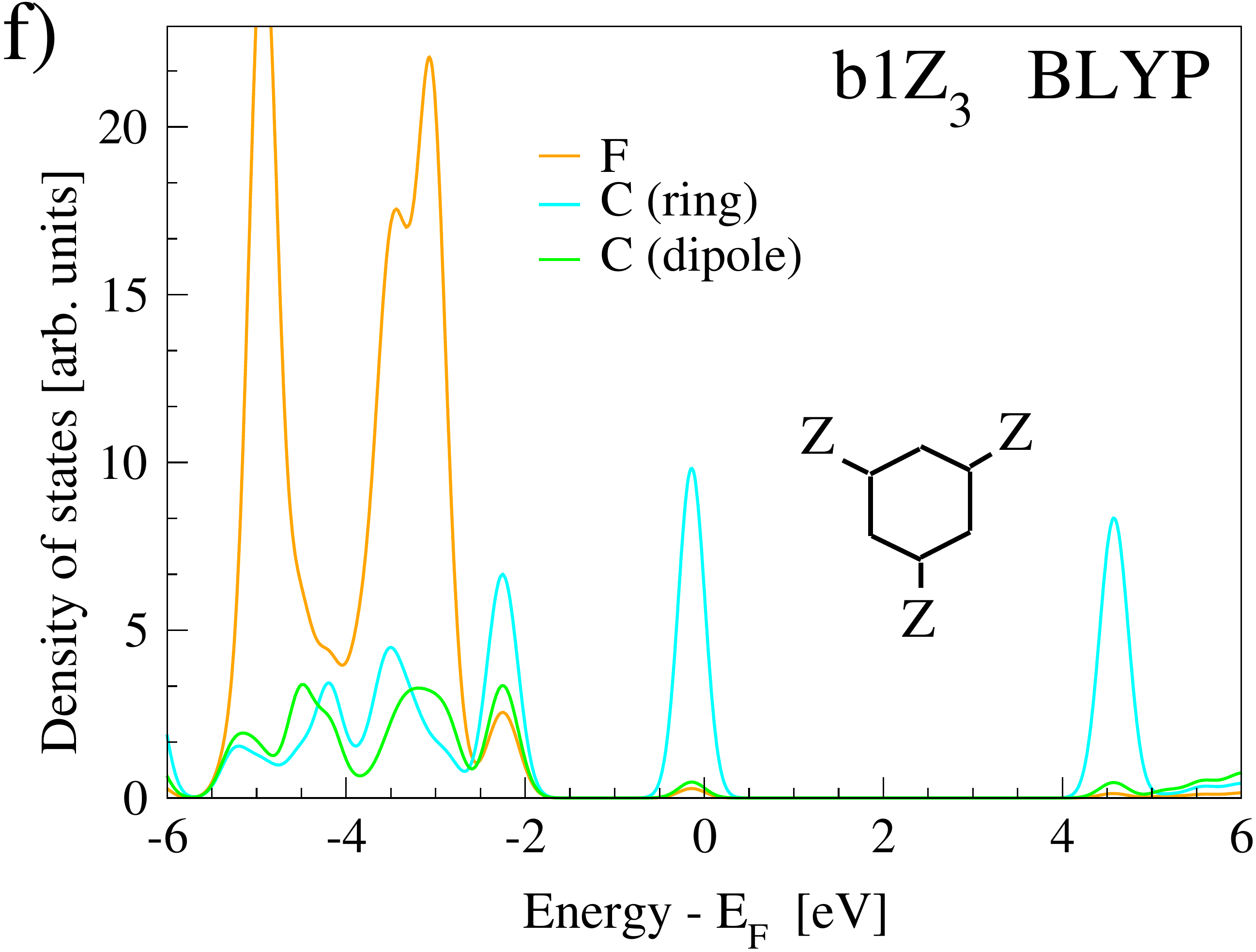} \\[0.3cm]

\includegraphics[scale=0.17]{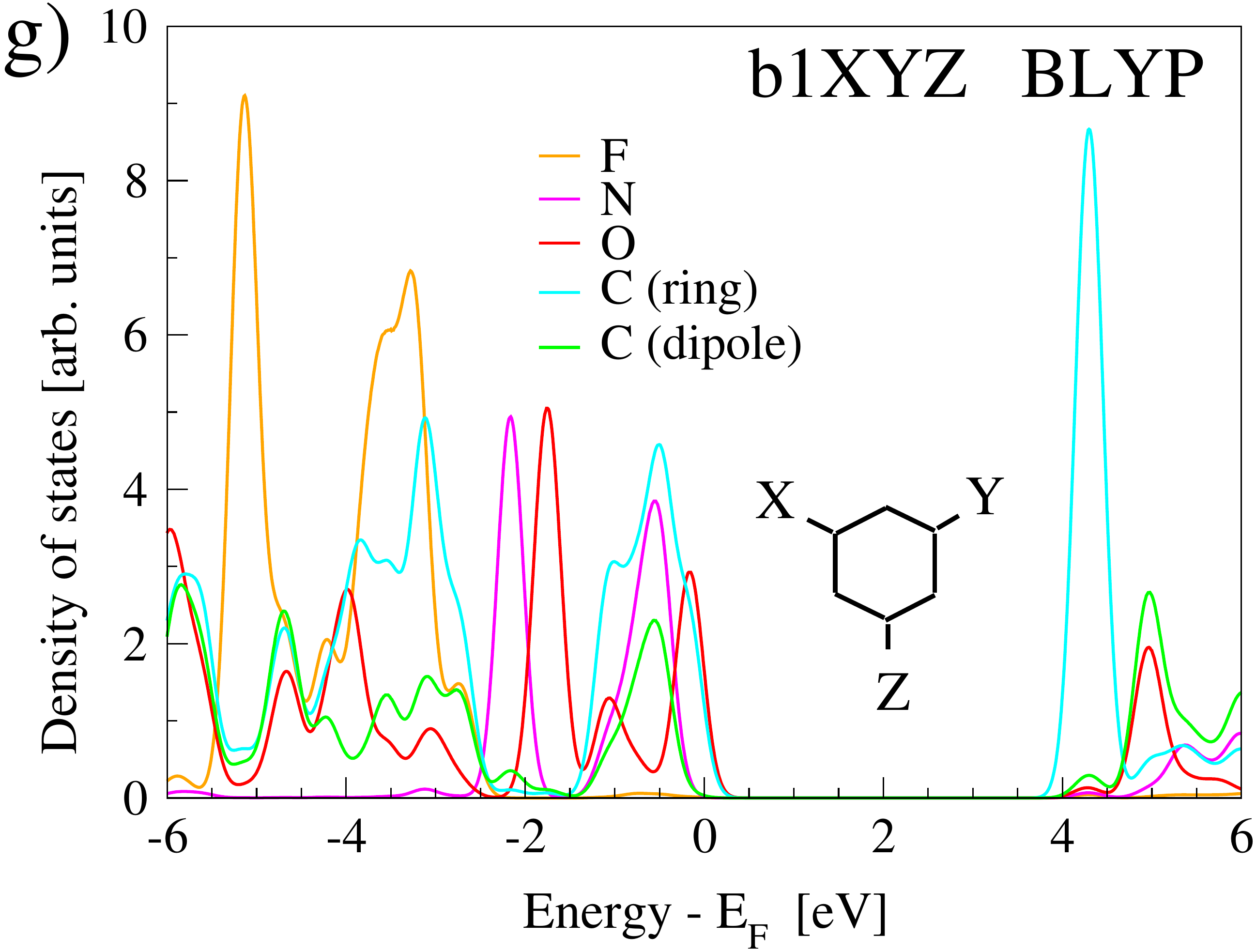} \hspace{0.2cm} &
\includegraphics[scale=0.17]{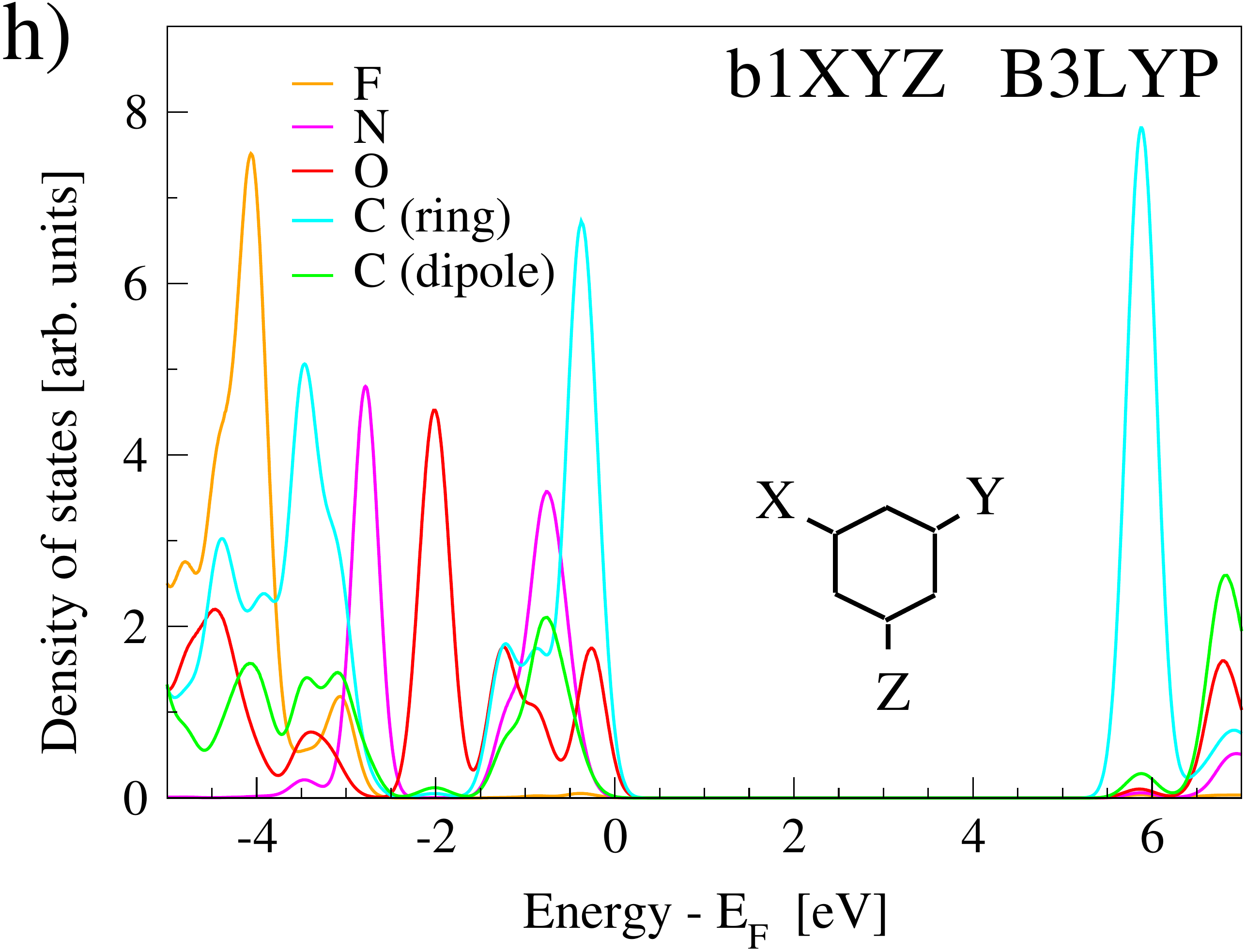} \hspace{0.2cm} & 
\includegraphics[scale=0.17]{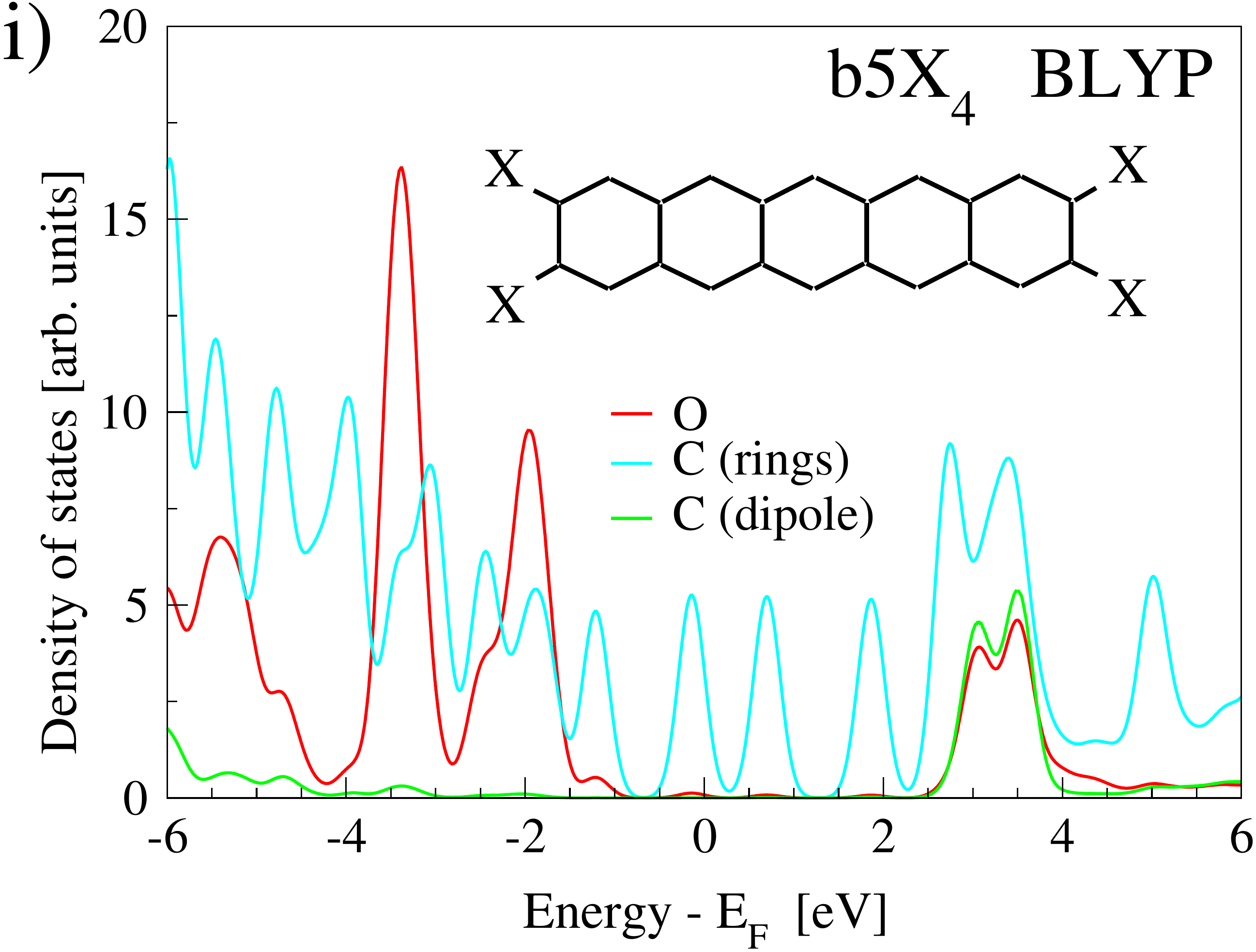} \\[0.3cm]

\includegraphics[scale=0.17]{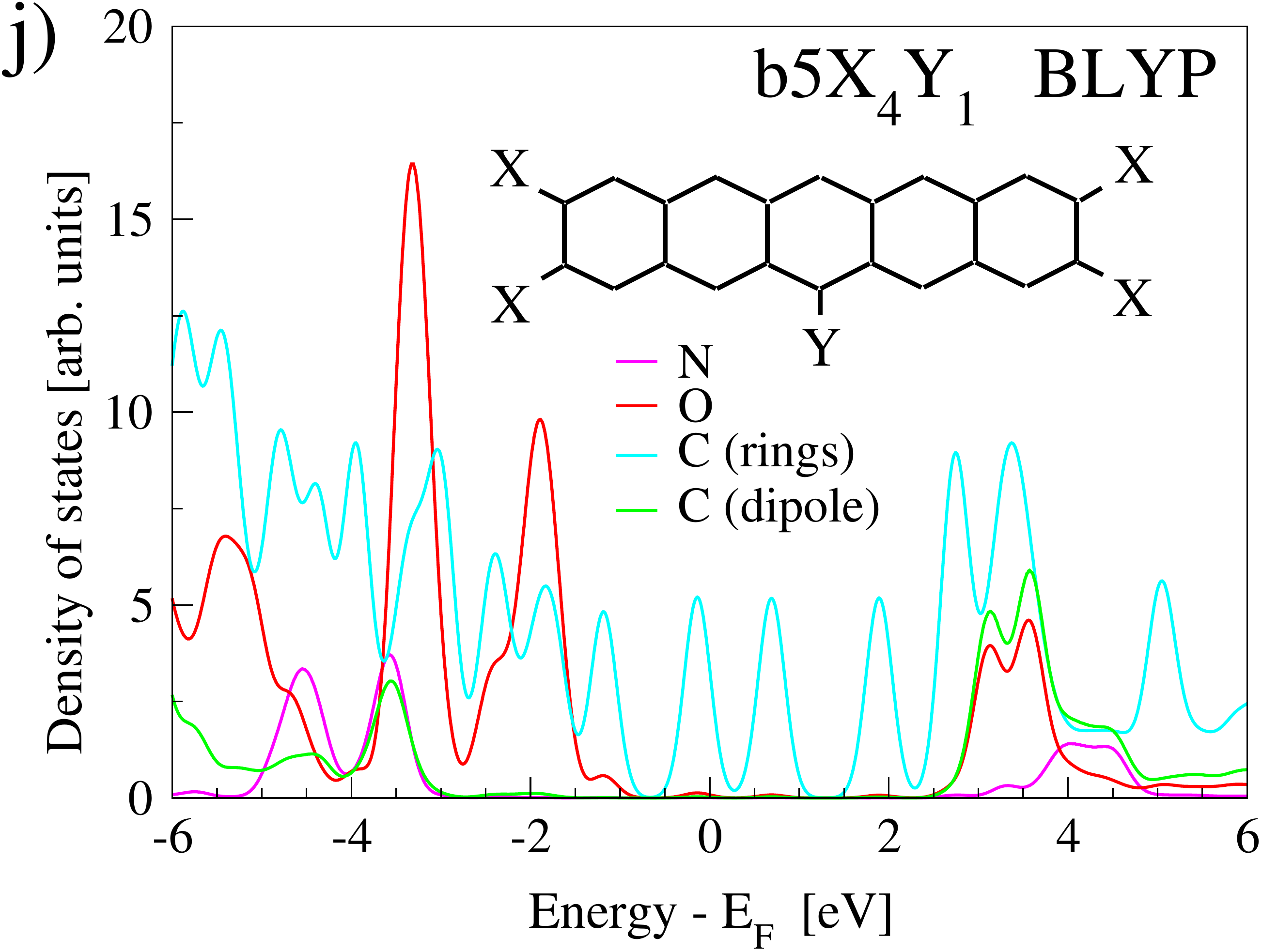} \hspace{0.2cm} &
\includegraphics[scale=0.17]{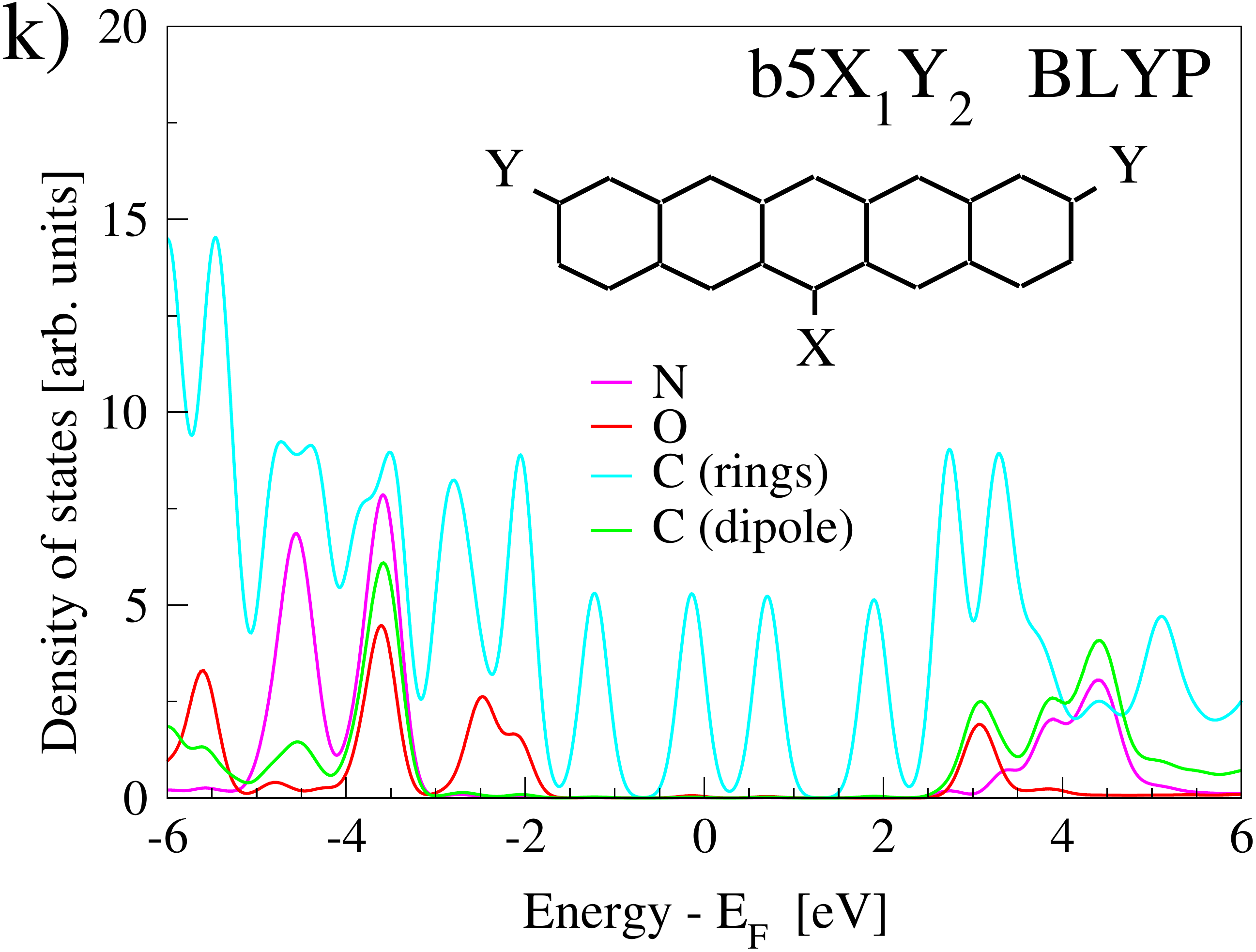} \hspace{0.2cm} &
\includegraphics[scale=0.17]{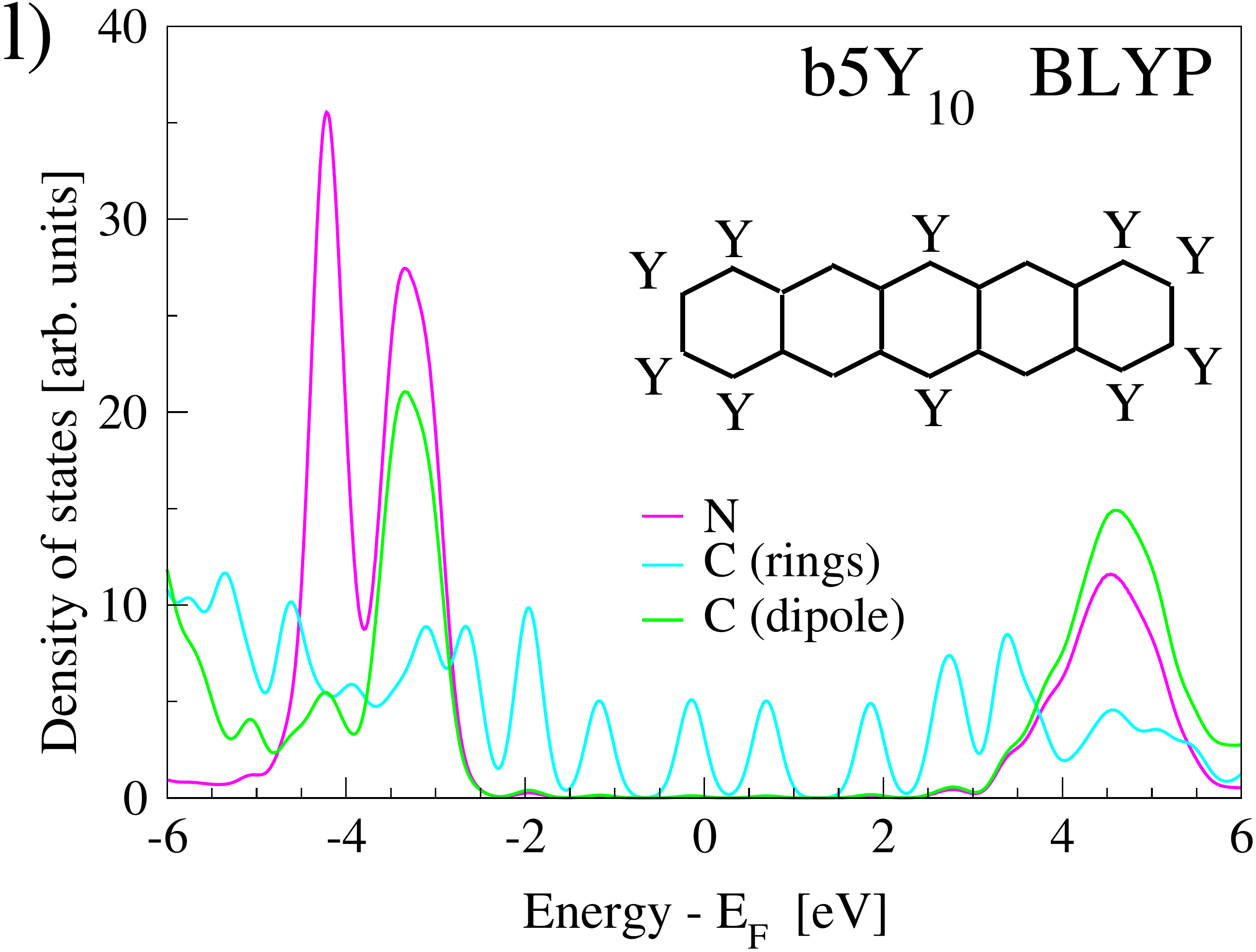} \\[0.3cm]

\includegraphics[scale=0.17]{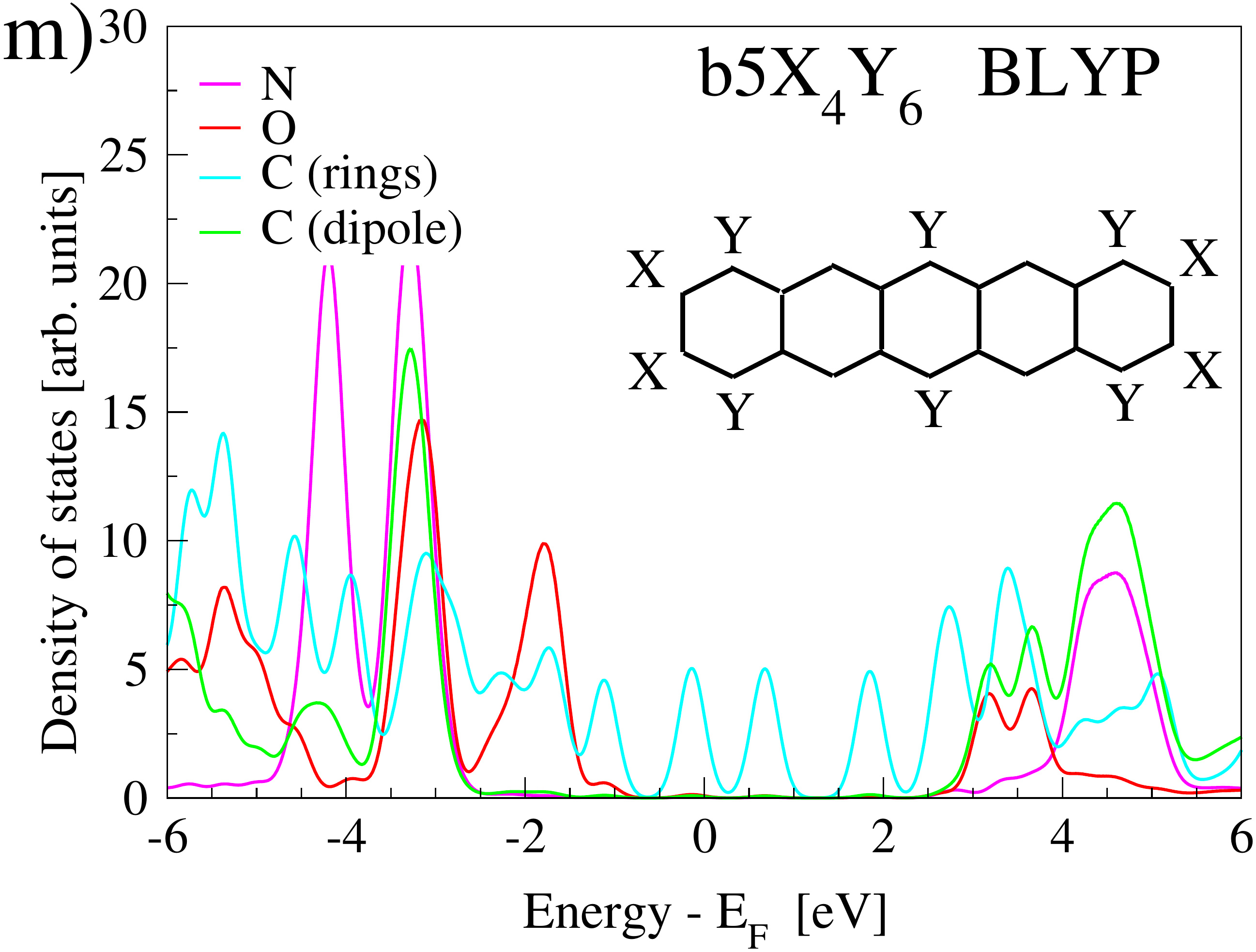} \hspace{0.2cm} &
\includegraphics[scale=0.17]{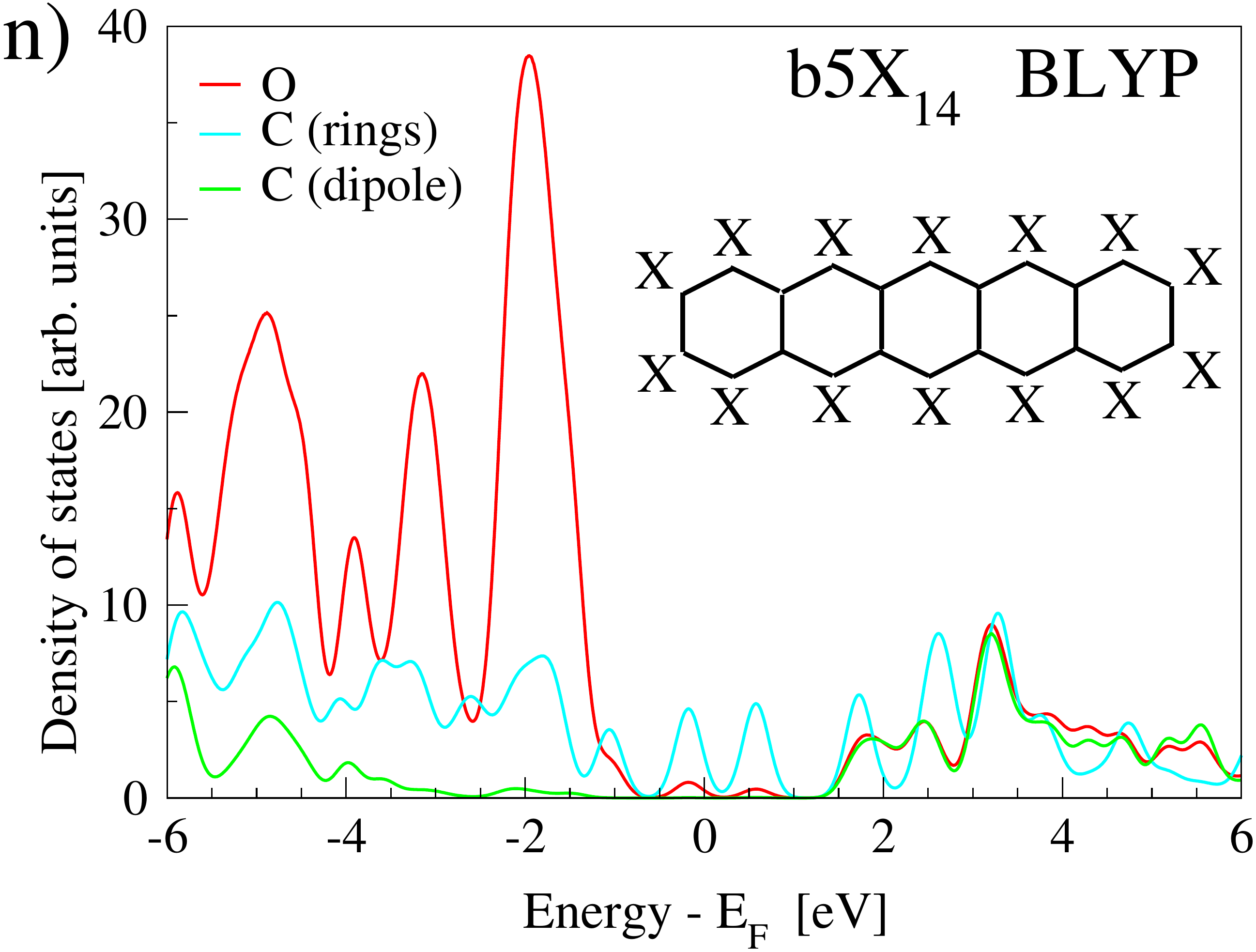} \hspace{0.2cm} &
\includegraphics[scale=0.17]{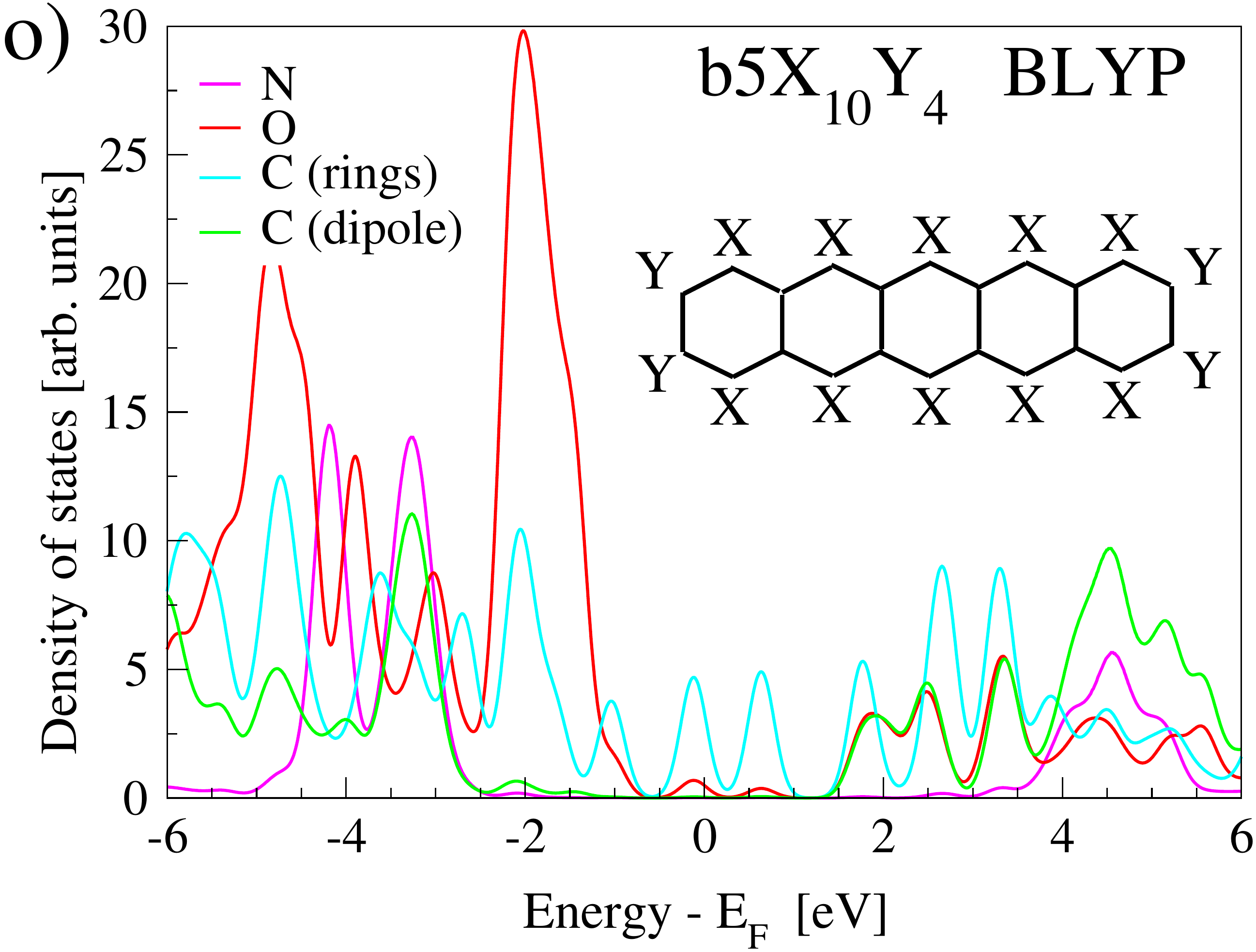}
\end{tabular}
\caption{Density of states (DOS) projected at the atoms of molecules; 
the systems and methods are denoted within the panels; 
$bn$ means a chain of $n$ benzene rings, X=COOH, Y=CH$_2$CN, Z=CH$_2$CF$_3$.}
\end{figure*}

In the previous work \cite{go2}, we showed that one of the systems studied here, 
namely a wire of b1(COOH,CH$_2$CN)$_3$ molecules, possesses very localized state 
at the conduction band minimum (CBM) and the valence band top (VBT).
Under the applied voltage, the electrons moved from the
central aromatic ring of the molecule to the neighboring ring, while the holes hopped
between the dipole groups. This property might reduce a recombination of carriers and
is very desired for solar cell devices \cite{perovskite}. 
Therefore, we characterize the states around the band gap for the wires composed of
some of the $\pi$-stacked molecules, studied in this work for their absorption energy.
Plots of the projected density of states (PDOS) for wires of benzene and pentacene decorated 
with various dipole groups are collected in Fig. 3.
The choice for the smallest molecule (benzene) is motivated by a need 
of making a comparison (of the results for the decorated benzene) to our previous studies
[7], using the hybrid-DFT method. While the pentacene molecule has been chosen, since it possesses
the gap in the sunlight range - and in the same time varying a type and an arrangement 
of the dipole groups. We varied the type and the arrangement of the dipole groups attached 
to these molecules in order to check the transport properties. As one will see further, changing 
the mesogenic part of the molecule does not alter the conclusions.

All results are obtained with the BLYP functional, except 
for only one case \-- namely b1(COOH,CH$_2$CN,CH$_2$CF$_3$) \--
for which the PDOS is calculated also with the B3LYP method. 
Figs. 3(g) and 3(h) compare these two methods and lead to the following conclusions: 
1) The orders of the energy bands are the same \-- specifically, the O-projected states are close to
the Fermi level, and N-projected DOS is deeper in the energy, 
while the C-ring PDOS is between the O- and N-projected states.
2) The only difference is for the width of the energy band localized at the C-ring \--
it is more narrow in the B3LYP case and leads to the higher value of the PDOS
at the edge of the valence band. Since using the more compuationally expensive method 
does not change the conclusions, we continue with the DFT approach for the PDOS analysis.

\begin{figure*}[b]
\begin{tabular}{cccc}
\includegraphics[scale=0.17]{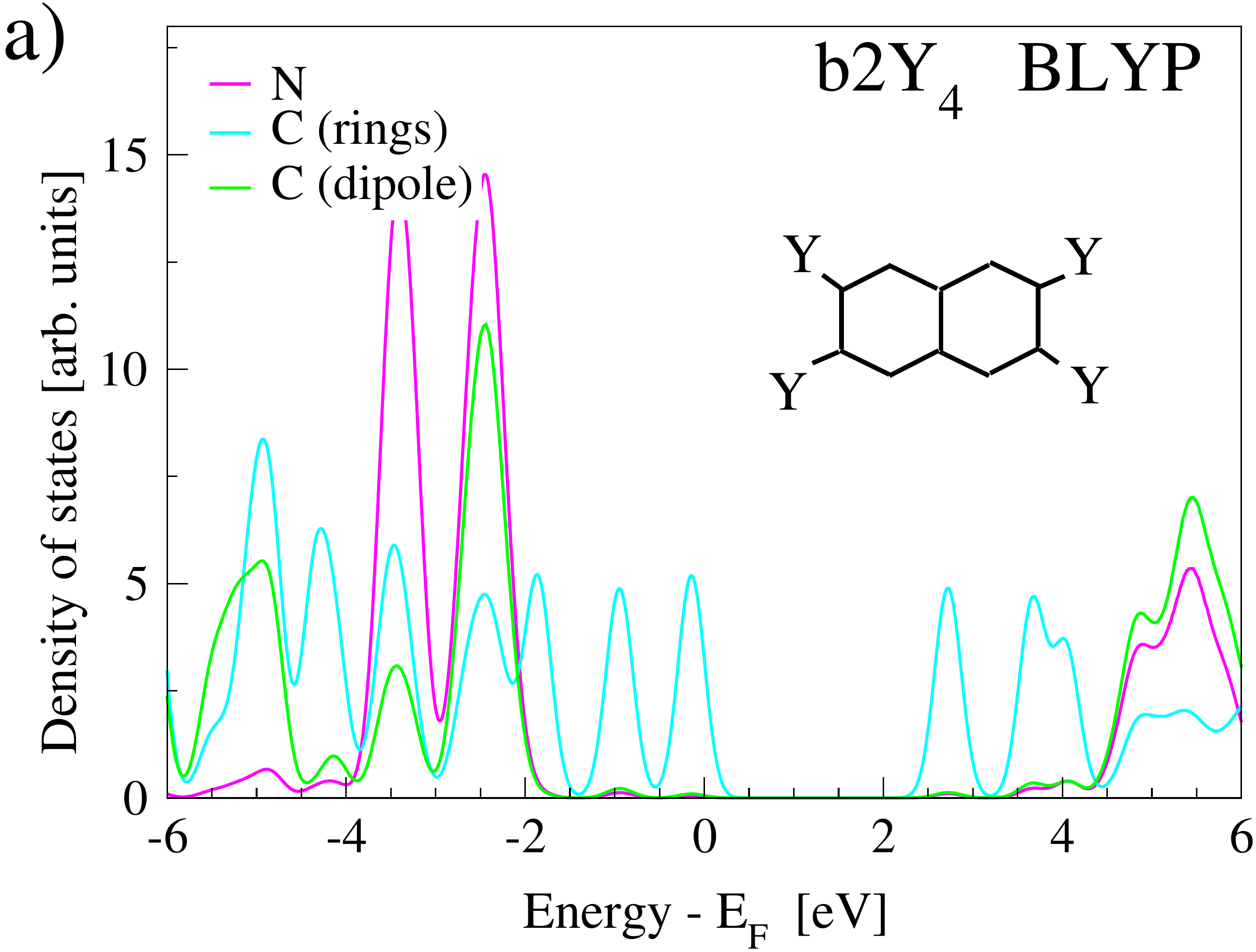} \hspace{0.2cm} &
\includegraphics[scale=0.17]{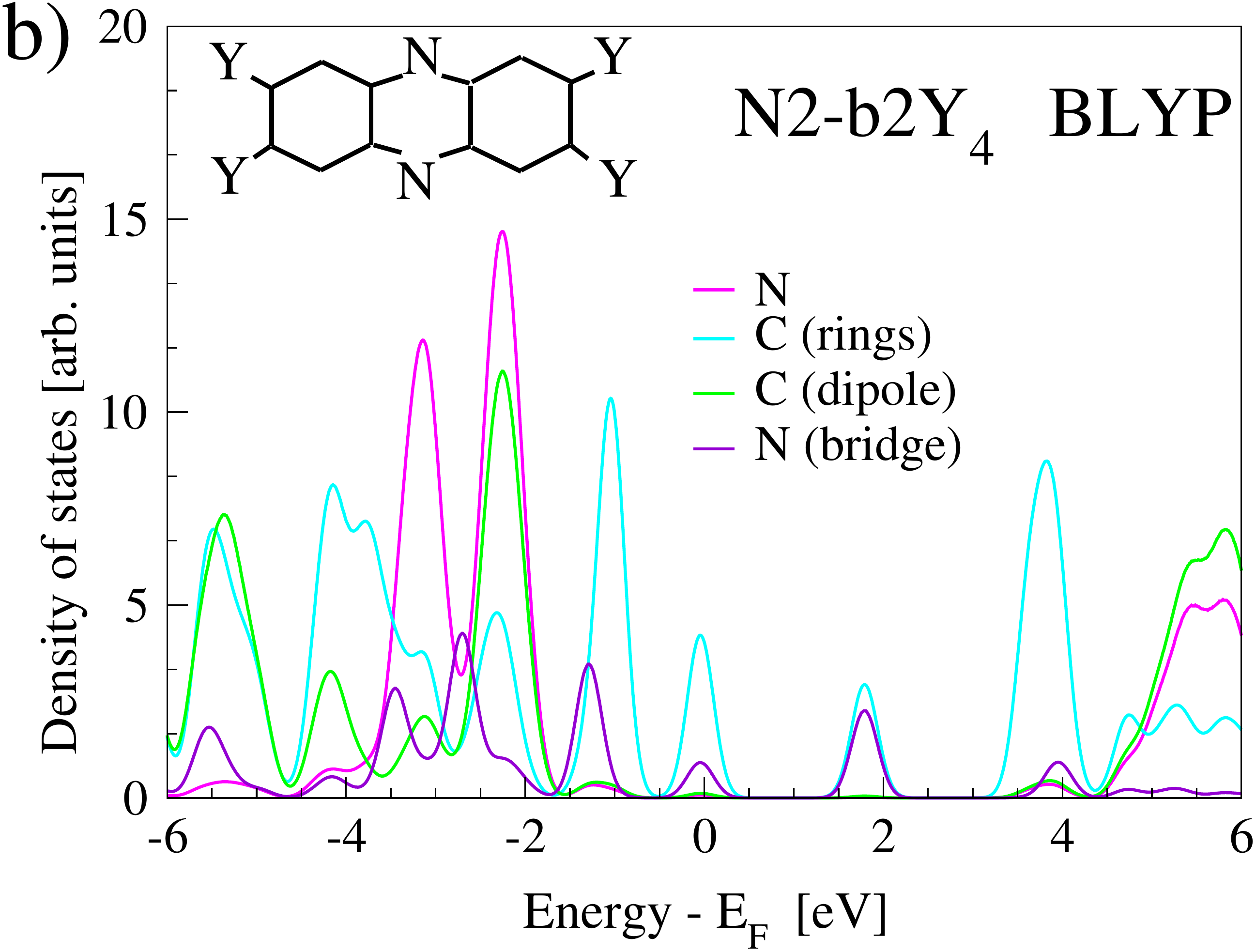} \hspace{0.2cm} &
\includegraphics[scale=0.17]{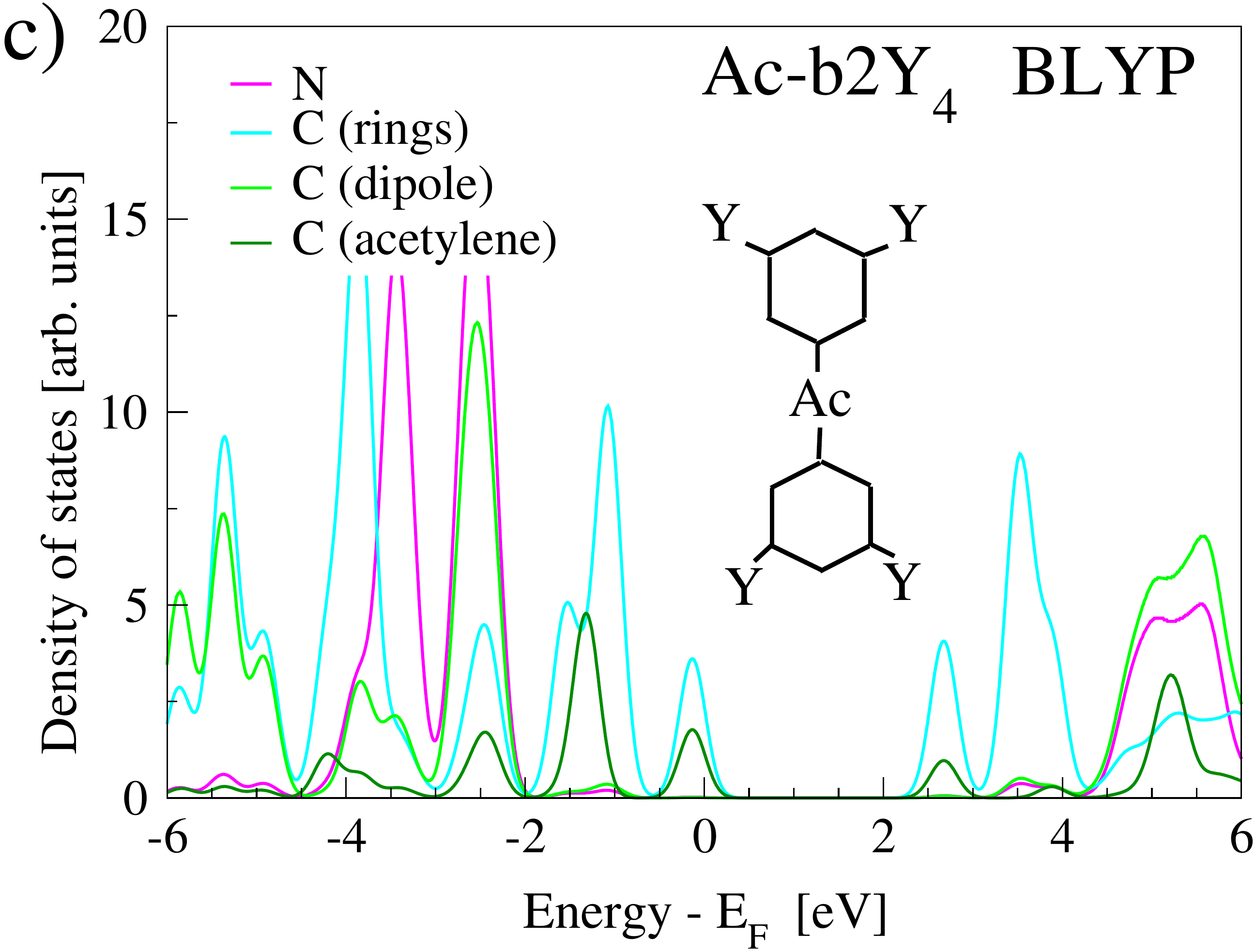}  \\[0.3cm]

\includegraphics[scale=0.17]{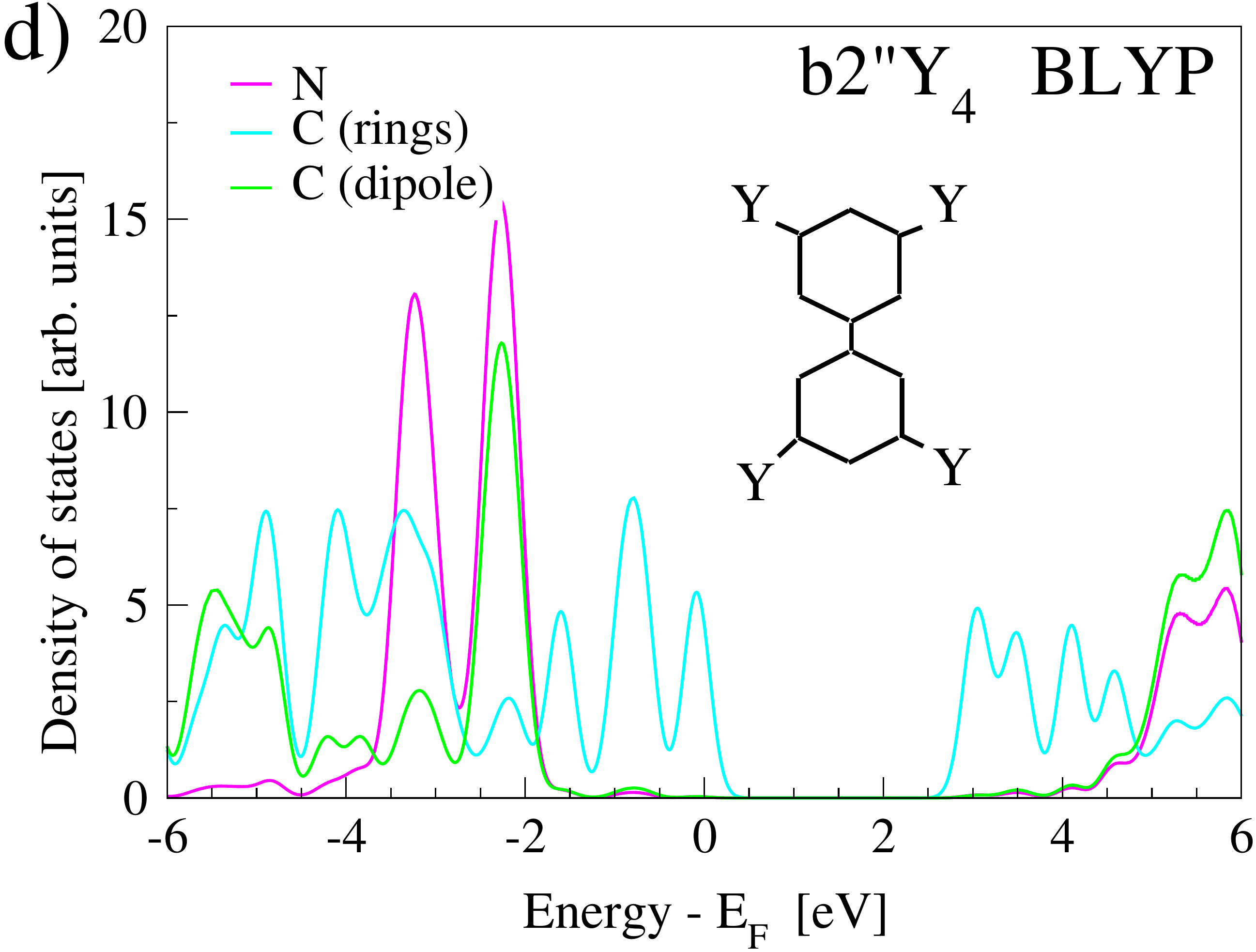} \hspace{0.2cm} &
\includegraphics[scale=0.17]{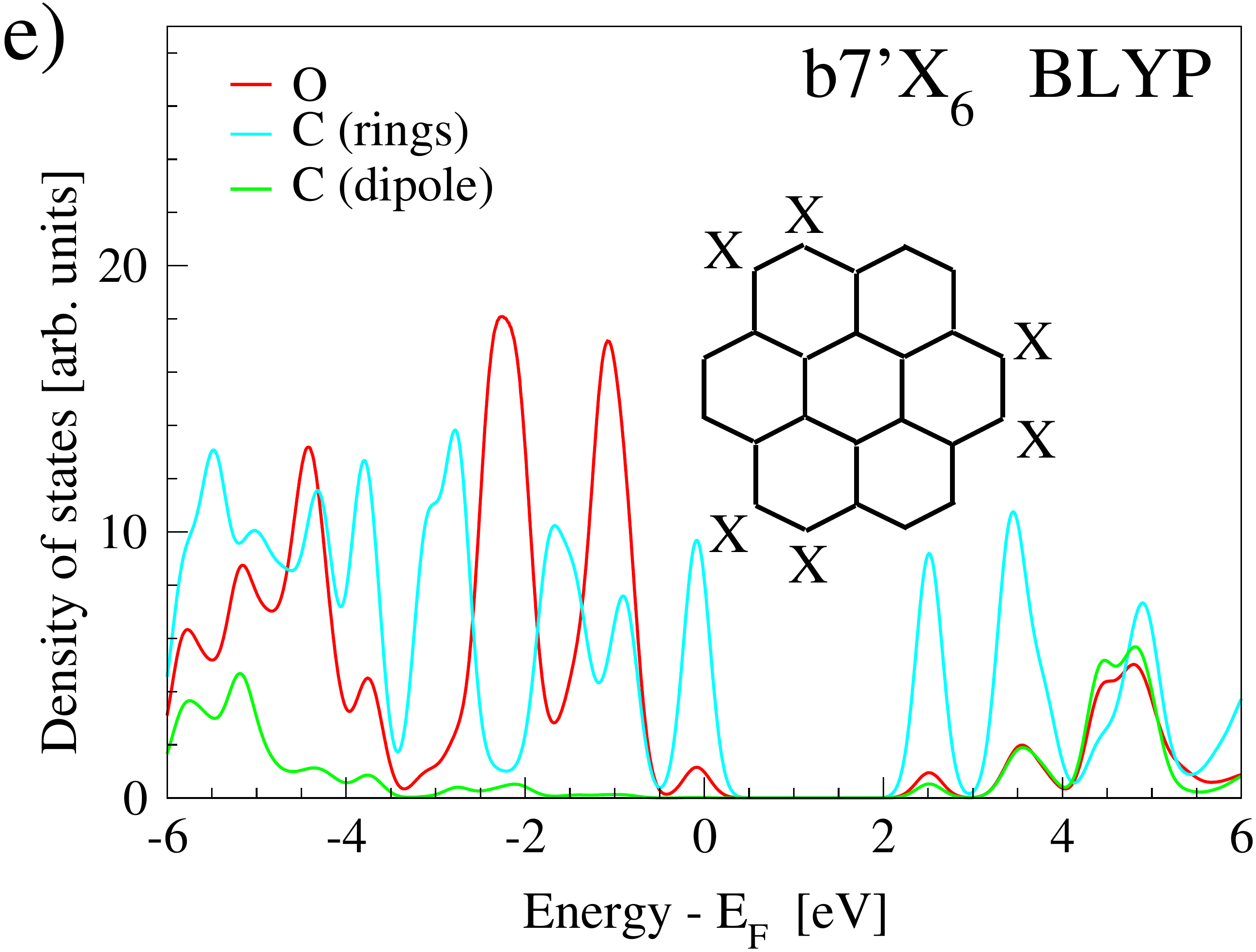}  \hspace{0.2cm} &
\includegraphics[scale=0.17]{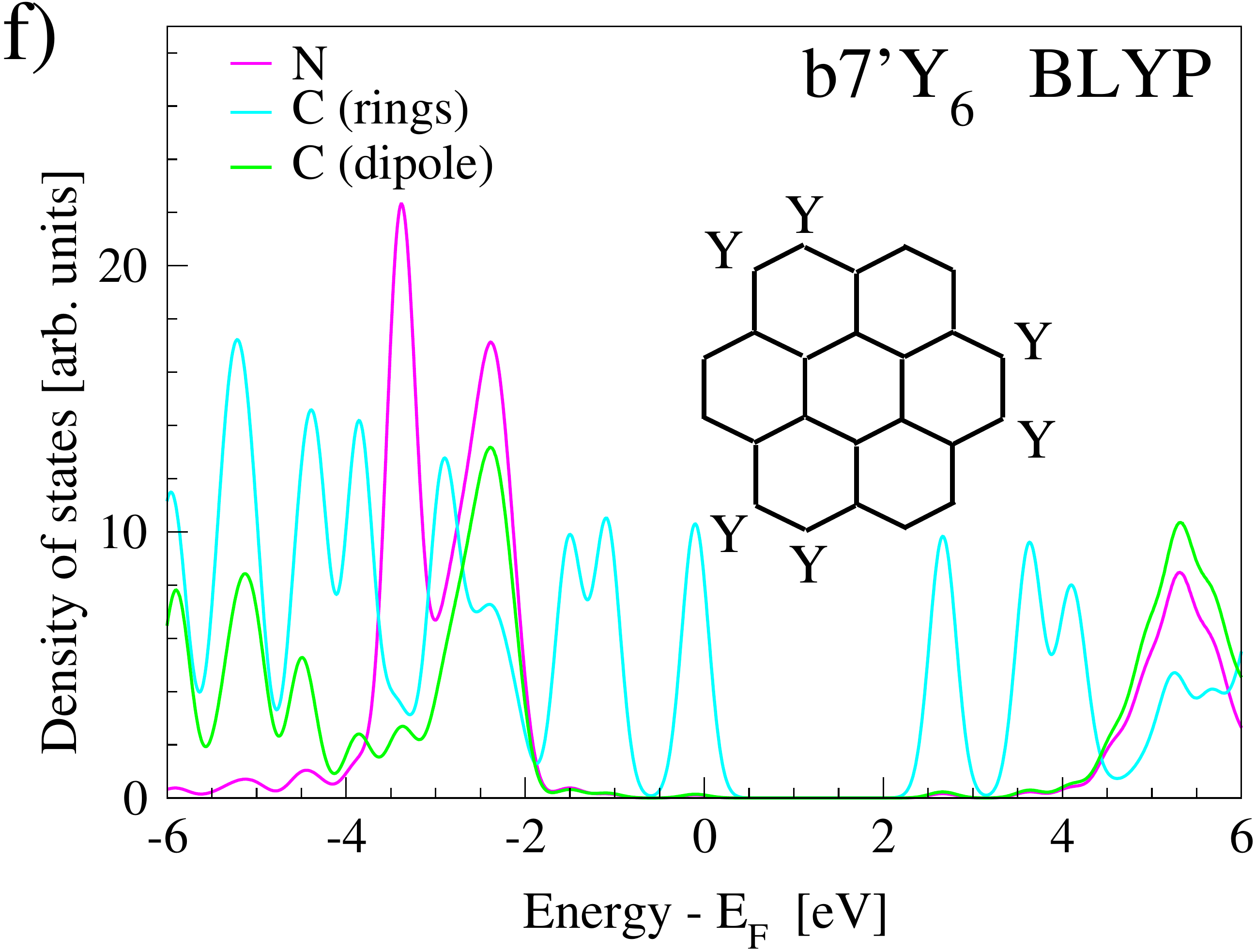}
\end{tabular}
\caption{Density of states (DOS) projected at the atoms of molecules;
the systems and methods are denoted within the panels;
$bn$ means a number of benzene rings, X=COOH, Y=CH$_2$CN, 
Ac is the acetylene bridge \--C$\equiv$C\--. The "prime" and "bis"
denote different ways of connecting the aromatic rings or shapes of 
the aromatic molecular cores. }
\end{figure*}

When the COOH or CH$_2$CN dipole groups are attached to a single benzene ring, the highest
occupied levels are composed of the states localized at dipoles and partially on the central aromatic ring,
while the lowest unoccupied states are built mainly of the C-ring localized orbitals. 
The oxygen orbitals are closer to the Fermi level than states of the N origin.
The fluorine states are the deepest in the energy of all studied dipole groups.
Moreover, if only CH$_2$CF$_3$ groups are attached to benzene then both holes and electrons
are predicted to move through the central ring, which is a very unwanted situation.
Summarizing results in Figs. 3(a)-(h): 
the best space separations of the carrier paths are for CH$_2$CN and COOH.
Combinations of these groups work as well. The COOH group is used as a connecting
part for the planar structures, as studied in the previous work \cite{go1}. 
It has been demonstrated \cite{adsor}, that using the COOH group for a deposition of 
the optical material 
at the transparent electrodes in the solar cell devices one can obtain 
the highest photovoltaic conversion of all experimentally tested contacts.
Thus, only combinations of the COOH and CH$_2$CN dipoles attached to pentacene
molecules are studied further, because they represent the group of molecules
which have the energy gaps falling within a considered range of the Sun spectrum.

In Figs. 3(i)-(o), for each studied case, the highest occupied states of the dipole-group origin
are separated from the Fermi level by at least one band.
It seems that the carriers separation, found for the decorated benzene 
in our previous studies \cite{go2}, may not work very well for larger molecules. 
However, the PDOS for the dipole localized states is very high,
which might cause that the oscillator strength for the absorption 
is larger than that for the band positioned at the Fermi-level.
The calculations of the optical properties by means
of the GW+BSE aproach are necessary for a full insight. On the other hand, 
when one builds the planar systems with $\pi$-stacking, and adds the electrodes, and applies the voltage 
\-- which leads to the Stark shift of the energy levels \-- then
the hole transport between the dipole groups omitting the central aromatic frame might be plausible.

Additionally, we checked a few more possible connections of two aromatic rings, through: 
(i) the acetylene (-Ac-), or (ii) nitrogen (-N-) bridges, or (iii) -C-C- bond.
The PDOS of naphthalene and bi-phenyl with four CH$_2$CN groups, and similar systems with the above bridges,  
are presented in Figs 4(a)-(d). 
There is only a little improvement of the positioning of the dipole groups in 
the PDOS, and it is when the -N- bridge is used, with respect to the other possibilities. 
This means that after an addition of COOH, the light holes will move between the carboxyl groups
and the heavy holes between the CH$_2$CN dipoles, if many terminal groups are attached to the 
edges. Figs 4(e)-(f) display the projected DOS for the decorated coronene molecule. 
They confirm earlier findings for the linear aromatic molecules. 

As promised earlier in this work,
we comment now on the two stacking effects mentioned in the preceding subsection.  
The strength of the bandgap lowering due to stacking is related to the bands broadening close to
the Fermi level. This, in turn, is a function of a strength of the interaction between the neighboring
molecules, and of course the distance between them  
(the closest neighboring groups of atoms are the CH$_2$ moieties below the molecular plane
and the CN or CF$_3$ tops of the dipoles below). Since addition of the COOH group moves 
the levels of other dipoles down in the energy, 
the band broadening of these deeper states does not affect so much the energy gap.
Therefore, addition of COOH weakens the effect of stacking
on the gaps of molecules with CH$_2$CN or CH$_2$CF$_3$ dipoles.
In the same way, the B3LYP method \-- which lowers a contribution of the dipole-group projected DOS
at the Fermi level with respect to the BLYP \-- 
leads to a smaller stacking effect on the band gap. The latest effect is due to the fact that, 
the C-ring states of the neighboring molecules originate from the groups of atoms, which 
are more distant than the neighboring dipoles in the stack.

\subsection{Absorption spectra of three molecules in a series of the growing size of the mesogenic part}

\begin{figure*}[b]
\vspace{5mm}
\centerline{
\includegraphics[scale=0.35]{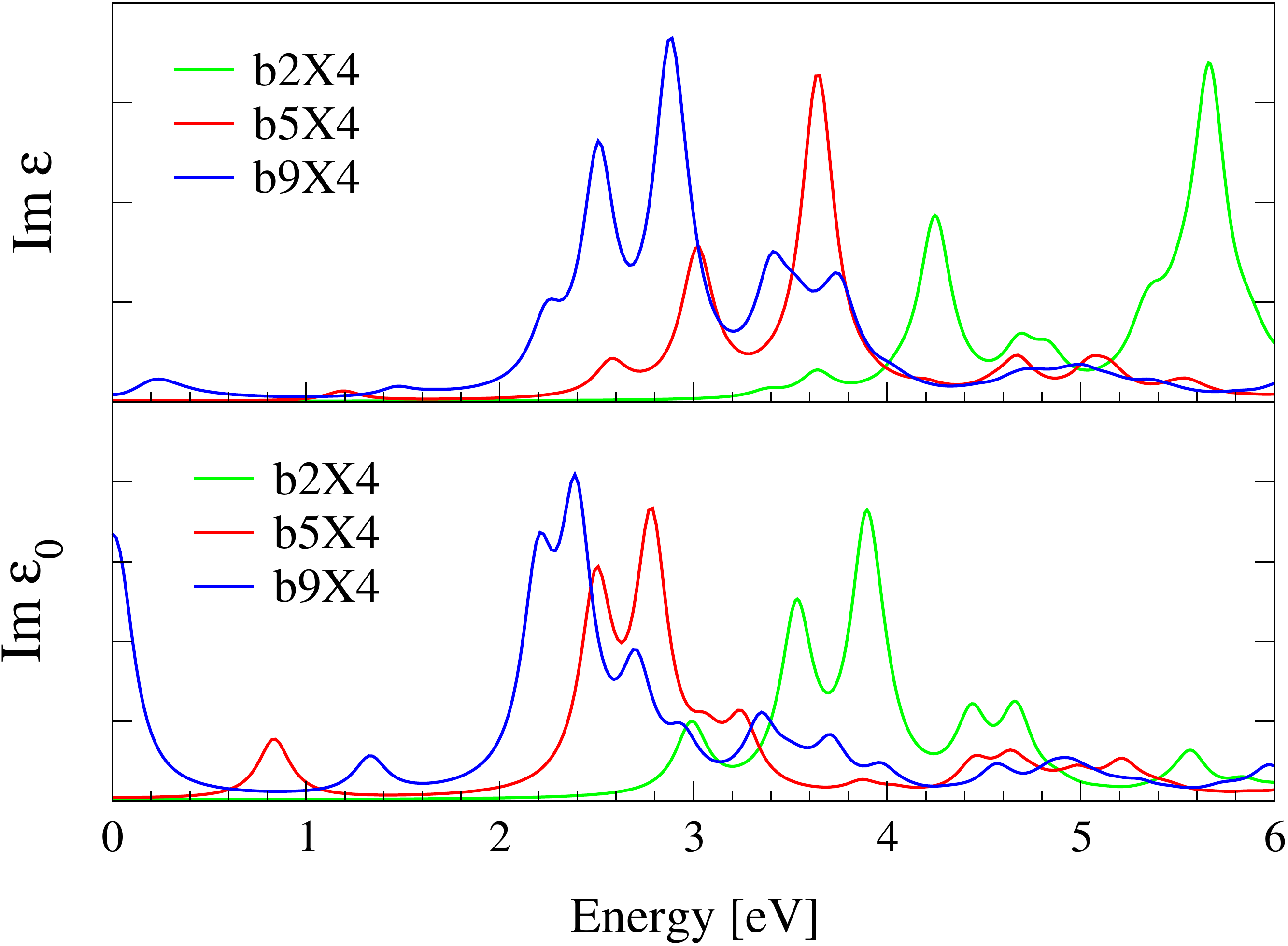} }
\vspace{5mm}
\caption{The imaginary part of the interacting (top panel) and noninteracting (bottom panel)
dielectric function for three chosen molecules, obtained with the Yambo code.} 
\label{b5}
\end{figure*}

At the end, we compare the theoretical absorption spectra for a series of three molecules
with growing number of benzene rings in a chain, i.e. 2, 5 and 9, 
but having the same number and type of the dipole groups, 
namely four COOH moieties attached on the both sides of the longer molecular axis, as in Fig. 1 
for b9(COOH)$_4$ (i.g. b9X4) and in Fig. 3(i) for b5X4.
These spectra are simulated with the Yambo code \cite{yambo}, using its possibility to calculate
the dielectric function. The response function was obtained on the random phase approximation level.
In Fig. 5, both the interacting ($Im\;\varepsilon$) and noninteracting ($Im\;\varepsilon_0$) 
dielectric functions show
the blue shift of the dominant absorption peaks with the growing molecular size.  

The imaginary part of the noninteracting dielectric functions have the absorption edges at the energies
which correspond to the DFT energy gaps \-- which are: 2.99 eV for b2X$_4$, 0.84 eV for b5X$_4$,  
and 0.04 eV for b9X$_4$. Measured absorbance is more similar with the interacting dielectric function,
which has the prominent peaks between 2 and 4 eV for b5X$_4$ and b9X$_4$, and between 4 and 6 eV
for b2X$_4$. The GW+BSE spectrum would be shifted up due
to the many-body effects and down due to the excitonic effects. Nevertheless, our main message that
it is possible to tune the optical spectra of the dipole decorated molecules in the $\pi$-stacks is
still valid. 

The second optimistic message is connected to the transport properties, which are
correlated with the oscillator strenth of the dipole optical transitions and 
creation of the electron-hole pairs.
As we see from the interacting dielectric function, the lowest absorption peak is much higher 
in the energy than the HOMO-LUMO gap.
This means that the optical transitions between the DOS peaks which are the most close 
to the Fermi level (on its occupied and unoccupied sides) are not allowed. These peaks were of the
same origin, namely they were localized at the central C-rings and not the dipole groups. 
Instead, the higher energetic positions of the first abroption peaks in these molecules
suggest that the allowed transitions are between the C-rings
and dipole groups. This is a wanted property, because the electron-hole pairs will be generated
on the spacially distant parts of the molecule. Thus, we retain the space separation of the charge
transport for electrons and holes again, for the larger molecules than benzene.   

\section{Conclusions}

Our aim is to tailor the band gap in the molecular $\pi$-stacks,
in order to propose systems \-- with ferroelectric properties investigated
earlier \cite{go1,go2} \-- for the solar cell applications.   
Tunning the optical properties is possible by varying a number the benzene rings
and a choice of the dipole groups when the molecules are small. 
While in cases of larger molecules, with longer aromatic chains, the band gaps are almost
independent on the terminal groups. On the other hand, a choice of the dipole groups and their
number are critical parameters for the atomic localization of the highest occupied and lowest
unoccupied states.
Therefore, the character of the photogenerated electron-hole pair is sensitive
to the chemical connections between the neighboring molecules in the stacks and this,
in turn, determines the transport properties \cite{go2}.  

Summarizing: the linear chains, between five and nine of the aromatic rings, are
good candites for building blocks of the organic nanostructes for photovoltaic
applications, when they are terminated with the COOH and CH$_2$CN groups. While
using the CH$_2$CF$_3$ groups does not give the desired properties.  
The dipole selection rules for the optical transitions retain the charge-paths selectivity,
which appears to be lost when looking just at the PDOS.

\section{Acknowledgement}

This work has been supported by The National Science Centre of Poland
(the Project No. 2013/11/B/ST3/04041).
Calculations have been performed in the Cyfronet Computer Centre using 
Prometheus computer which is a part of the PL-Grid Infrastructure, 
and by part in the Interdisciplinary Centre of
Mathematical and Computer Modeling (ICM). \\ 
 \\


\newpage

\setcounter{table}{0}
\setcounter{figure}{0}

\section{Supplementary information}


\begin{table*}[h]
\caption{Numerical data set for Fig. 2: the LUMO-HOMO energies (in eV) of the isolated molecules 
with variable number of benzene rings and three dipole groups: X=COOH, Y=CH$_2$CN, Z=CH$_2$CF$_3$.}

\begin{tabular}{cccccccccccccccccccc}
\hline \\[-0.2cm]

b2X$_4$ & b2X$_8$ & b5X$_4$ & b5Z$_6$  & b5X$_8$ & 
b5Y$_{10}$ & b5X$_{10}$ & b5X$_{14}$ & b9X$_4$  & b9X$_8$ \\

4.384 & 4.099 & 1.783 & 1.745 & 1.770 & 
1.766 & 1.748 & 1.677 & 0.820 & 0.814 \\

\\

b9Z$_{10}$ & b9Y$_{14}$ & b9X$_{14}$ & b9X$_{22}$ & b17X$_4$ & 
b17X$_8$ & b17Z$_{18}$ & b17X$_{22}$ & b17Y$_{22}$ & b17X$_{38}$ \\ 

0.808 & 0.821 & 0.810 & 0.778 & 0.568 & 
0.568 & 0.547 & 0.556 & 0.544 & 0.549 \\

\hline \\[-0.3cm]
\hline
\end{tabular}

\label{}

\end{table*}



\begin{figure*}[h]
\begin{tabular}{ccc}

b2X$_4$ & & b2X$_8$   \\

\includegraphics[scale=0.11,angle=0.0]{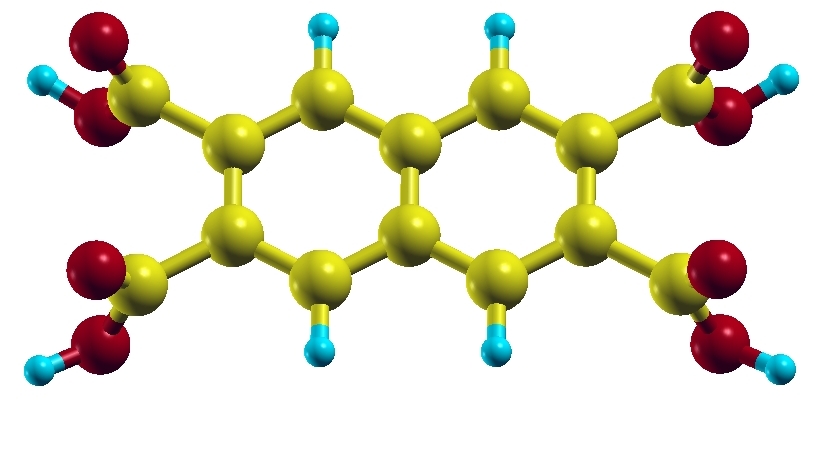} & 
\includegraphics[]{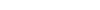} & 
\includegraphics[scale=0.10,angle=0.0]{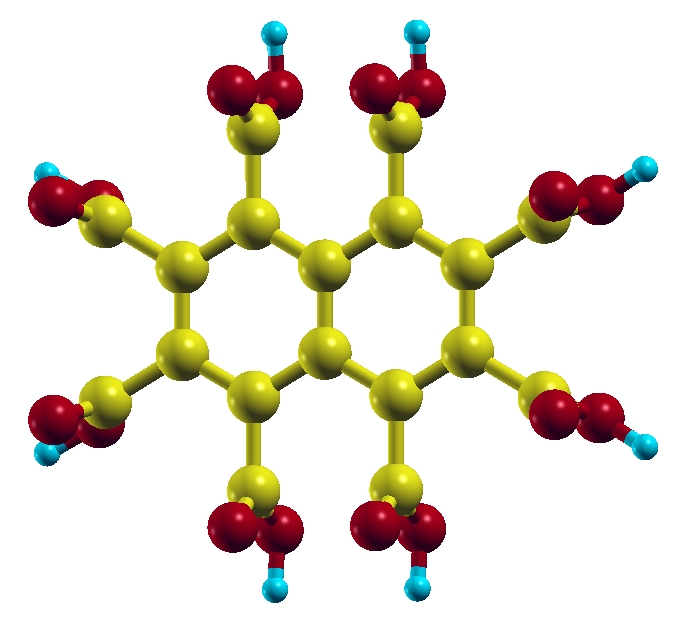} \\ [0.8cm]

\end{tabular}

\begin{tabular}{ccc}
 b5X$_4$ & b5Z$_6$ & b5X$_8$ \\

\includegraphics[scale=0.13,angle=-0.5]{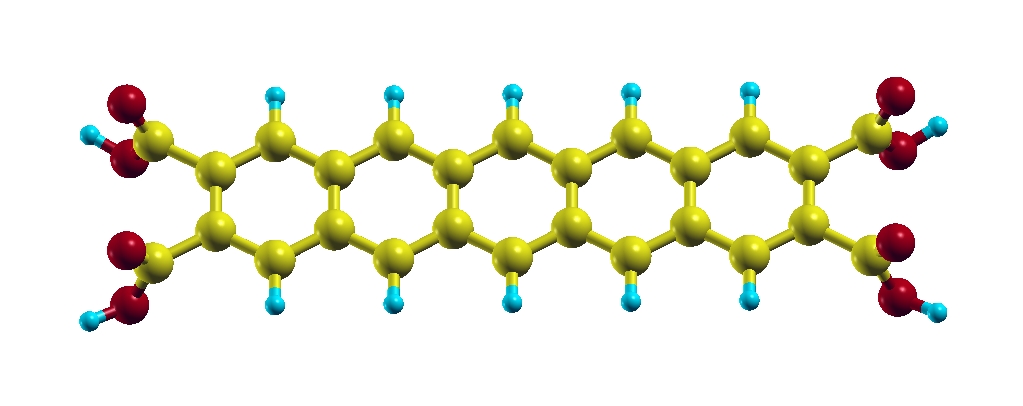} &
\includegraphics[scale=0.13,angle=-1.0]{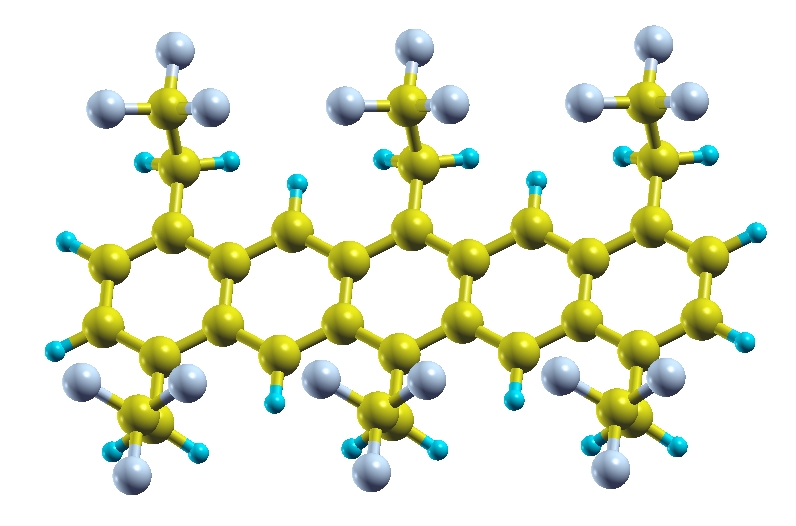} &
\includegraphics[scale=0.13,angle=-1.0]{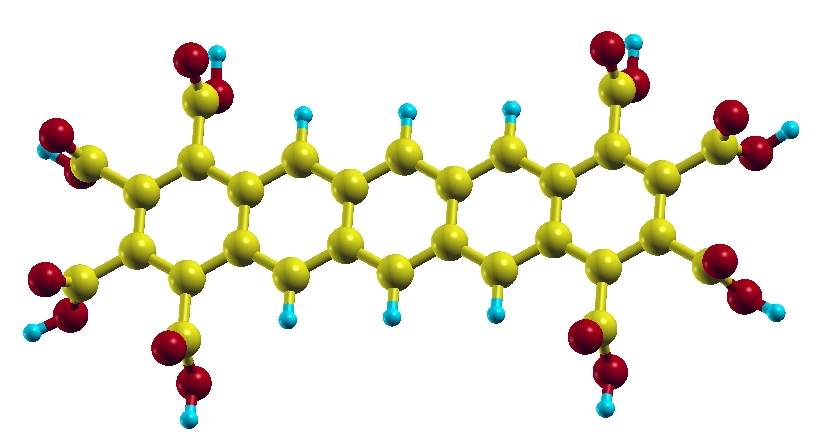} \\ [0.4cm]

b5Y$_{10}$ & b5X$_{10}$ & b5X$_{14}$ \\

\includegraphics[scale=0.13,angle=0.6]{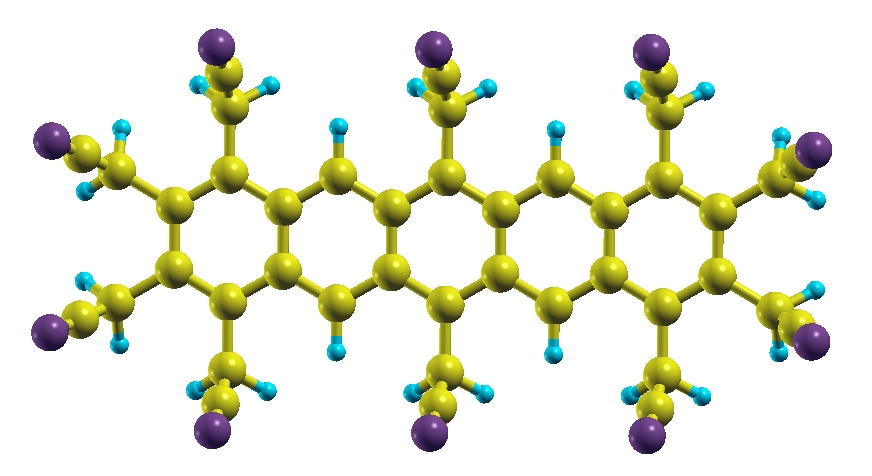} &
\includegraphics[scale=0.13,angle=-1.0]{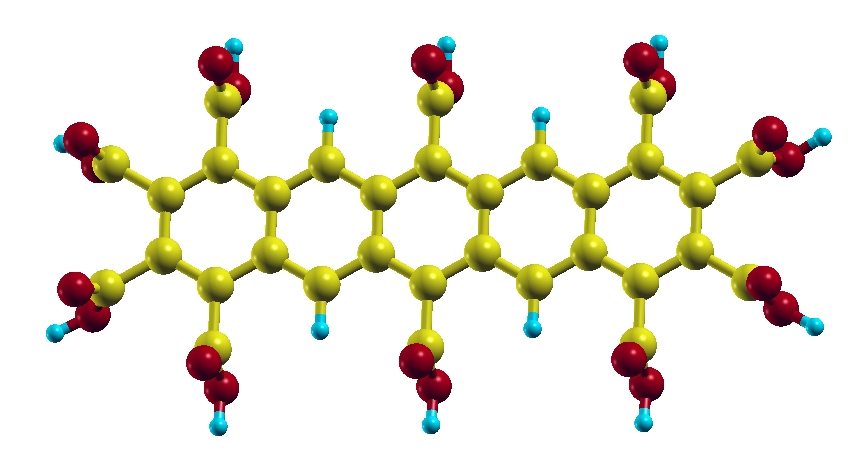} &
\includegraphics[scale=0.13,angle=-0.5]{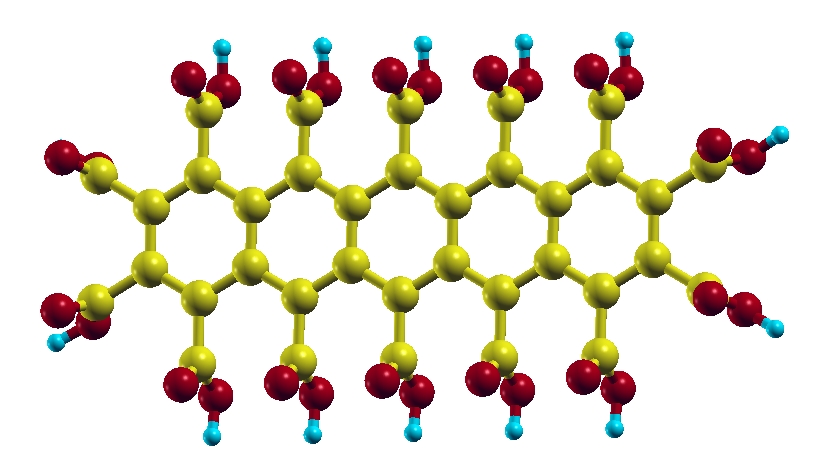} \\ [0.4cm]

b9X$_4$ &  b9X$_8$ & b9Z$_{10}$ \\

\includegraphics[scale=0.15,angle=0.0]{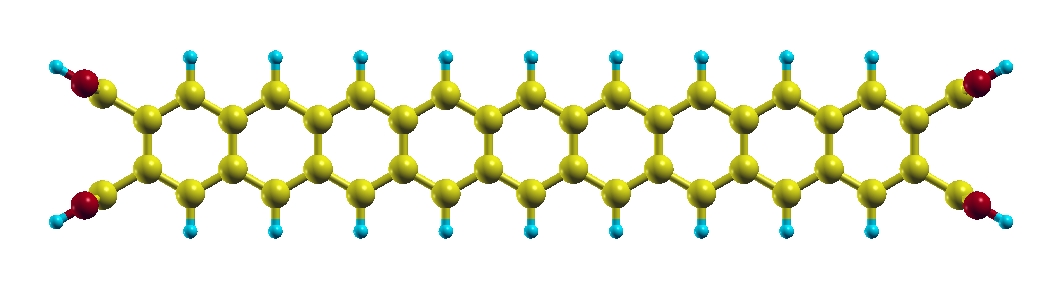} &
\includegraphics[scale=0.15,angle=-0.5]{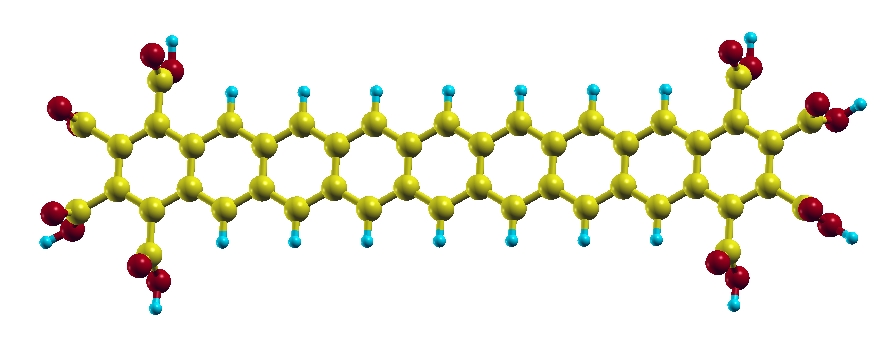} &
\includegraphics[scale=0.15,angle=-0.5]{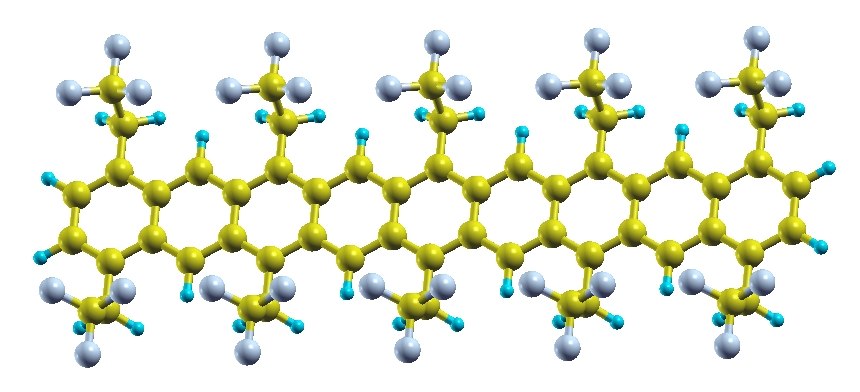} \\ [0.4cm]

b9Y$_{14}$ & b9X$_{14}$ & b9X$_{22}$ \\

\includegraphics[scale=0.15,angle=0.0]{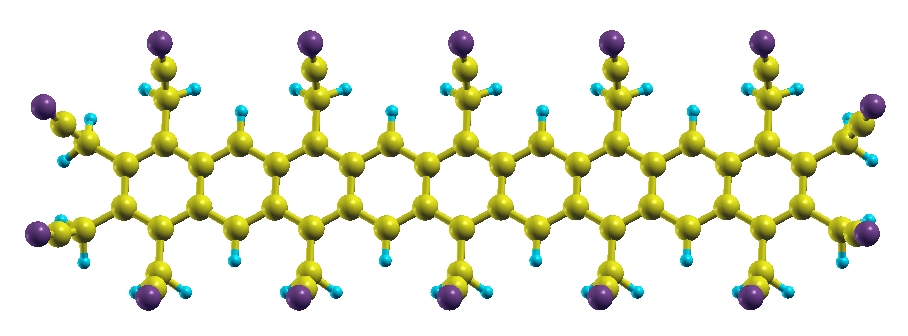} &
\includegraphics[scale=0.15,angle=0.0]{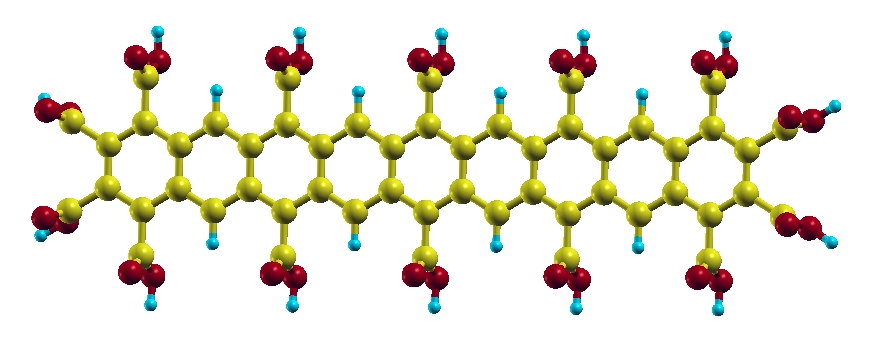} & 
\includegraphics[scale=0.15,angle=0.0]{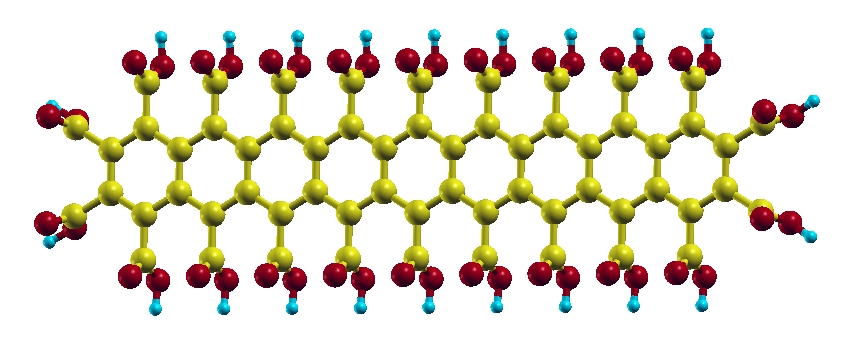} \\ [0.7cm]

\end{tabular}

\begin{tabular}{cc}

b17X$_4$ & b17X$_8$\\

\includegraphics[scale=0.20,angle=0.5]{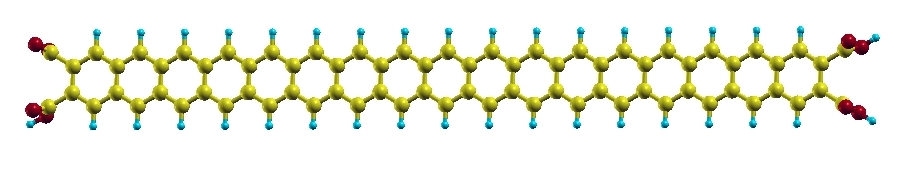} &
\includegraphics[scale=0.20,angle=-0.5]{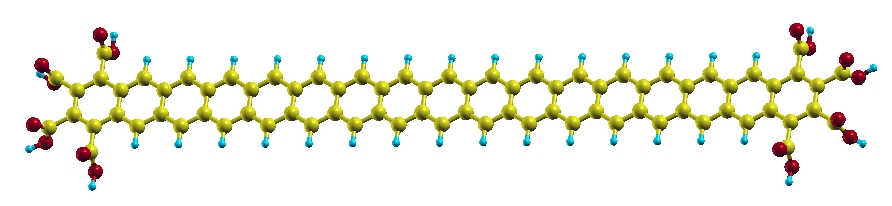} \\ [0.7cm]

b17Z$_{18}$ & b17X$_{22}$ \\

\includegraphics[scale=0.20,angle=-1.0]{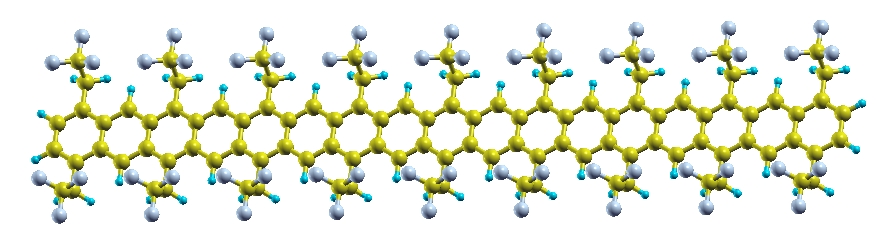} & 
\includegraphics[scale=0.20,angle=-0.5]{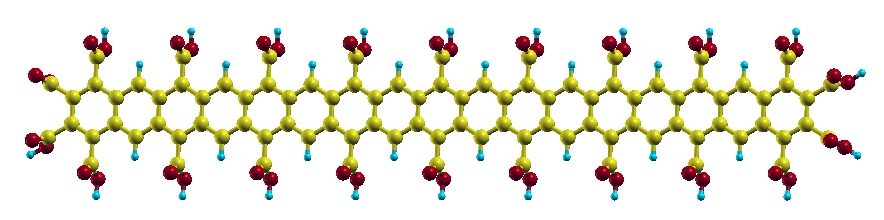} \\ [0.7cm]

b17Y$_{22}$ & b17X$_{38}$ \\

\includegraphics[scale=0.20,angle=0.0]{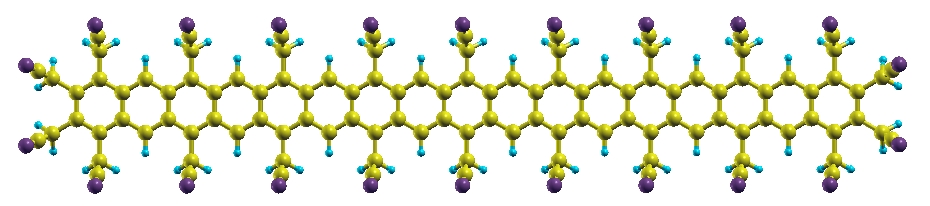} &
\includegraphics[scale=0.24,angle=0.6]{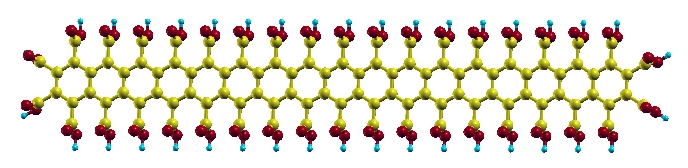} \\ [0.7cm]

\end{tabular}

\caption{Atomic structures of molecules for which the LUMO-HOMO gaps are presented in Figure 2.}
\end{figure*}

\begin{figure*}[h]
\begin{tabular}{cccc}
 b1(COOH)$_3$  &  b1(COOH)$_6$  &  b1(CH$_2$CN)$_3$  &  b1(COOH,CH$_2$CN)$_3$ \\

\includegraphics[scale=0.20,angle=0.0]{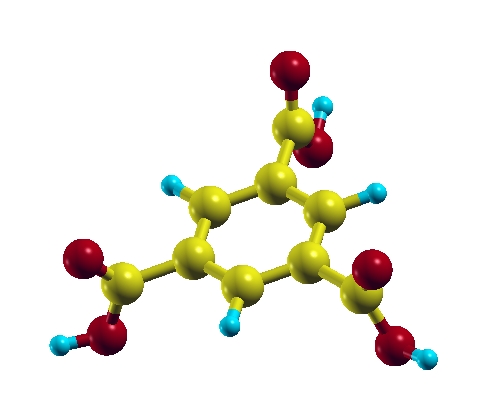}  & 
\includegraphics[scale=0.18,angle=0.0]{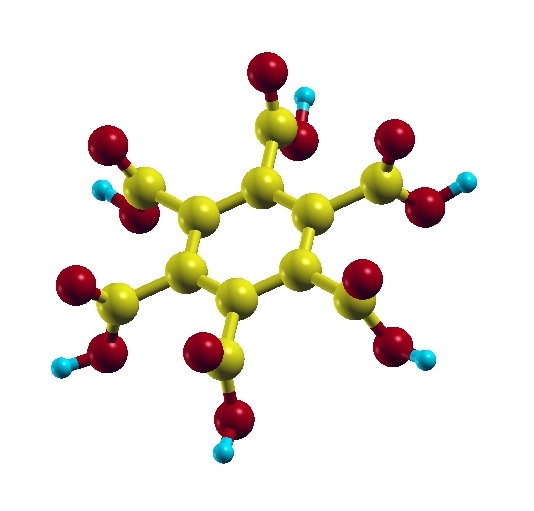} &
\includegraphics[scale=0.19,angle=0.0]{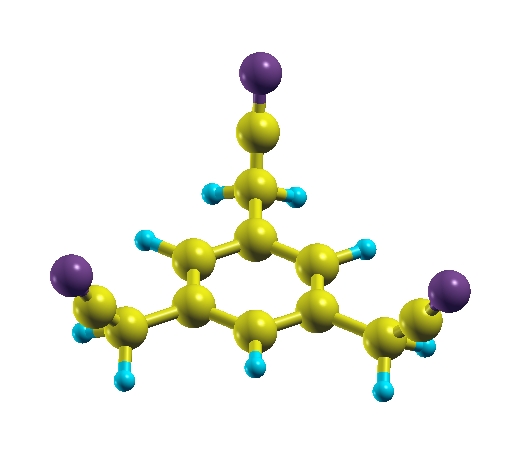} &
\includegraphics[scale=0.18,angle=0.0]{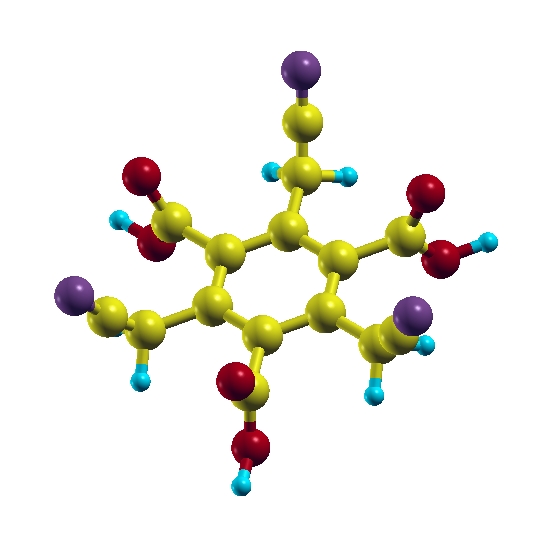} \\ [1.5cm]

b1(CH$_2$CF$_3$)$_3$  &  b1(COOH,CH$_2$CF$_3$)$_3$  &  \multicolumn{2}{c}{b1(COOH,CH$_2$CN,CH$_2$CF$_3$)}\\

\includegraphics[scale=0.23,angle=0.0]{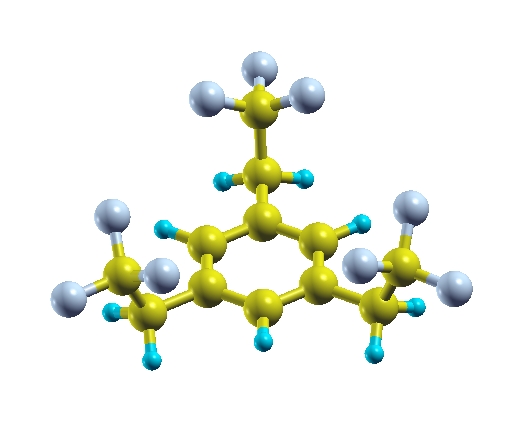} &
\includegraphics[scale=0.18,angle=0.0]{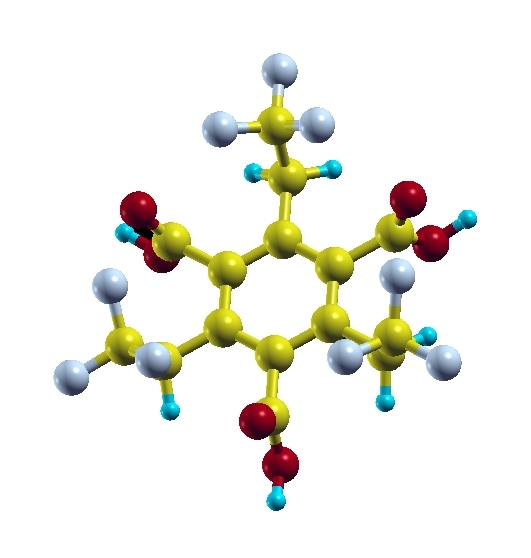} &
\multicolumn{2}{c}{\includegraphics[scale=0.22,angle=0.0]{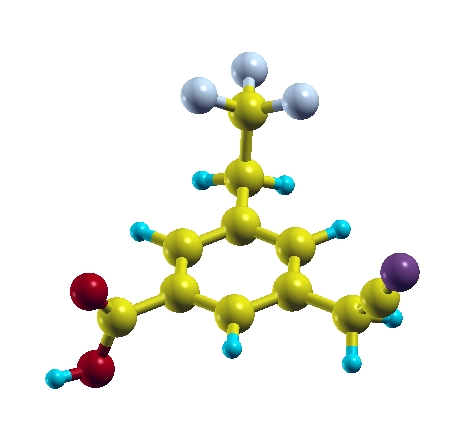}} \\ [1.5cm]

b5COOH$_4$  & b5(CH$_2$CN)$_{10}$  &  \multicolumn{2}{c}{b9(COOH)$_4$}  \\

\includegraphics[scale=0.14,angle=0.5]{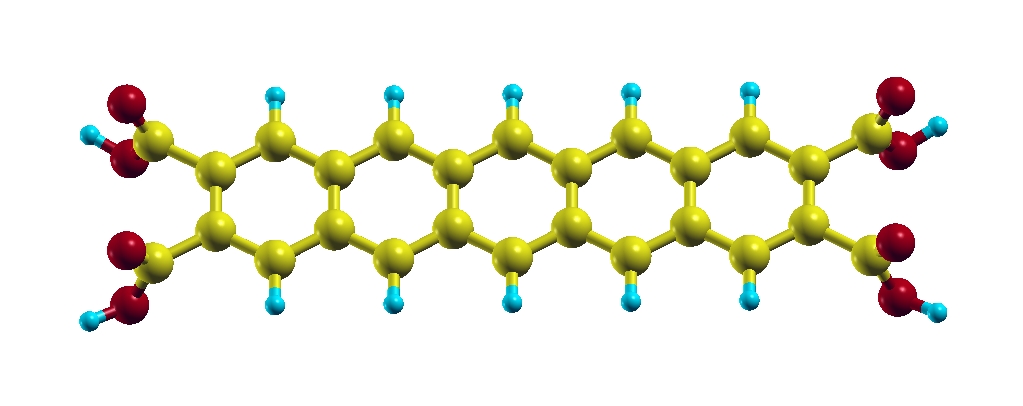} &
\includegraphics[scale=0.20,angle=-0.4]{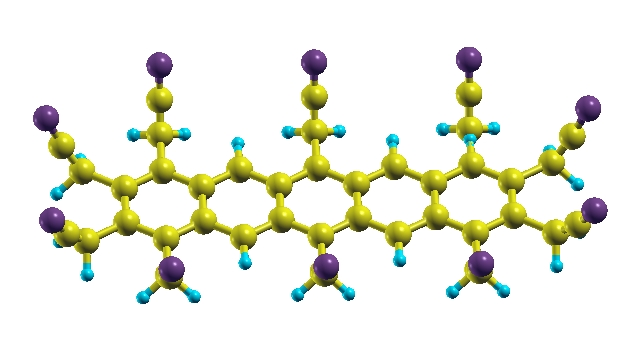} &
\multicolumn{2}{c}{\includegraphics[scale=0.18,angle=0.0]{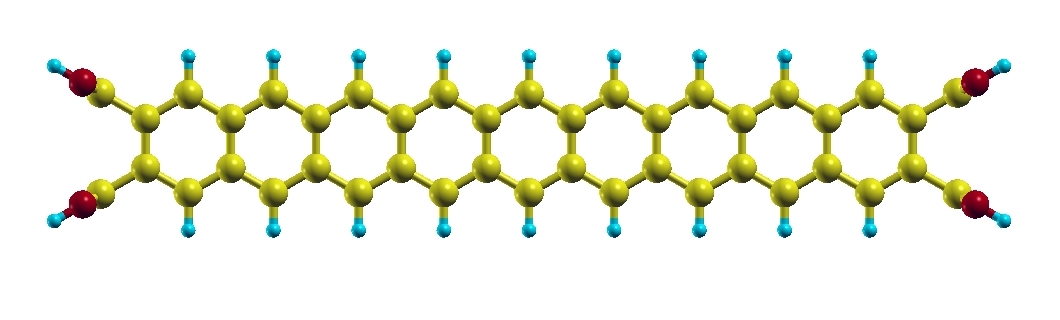}} \\ [1.5cm]

\multicolumn{4}{c}{b17(COOH)$_{8}$} \\

\multicolumn{4}{c}{\includegraphics[scale=0.4,angle=-0.8]{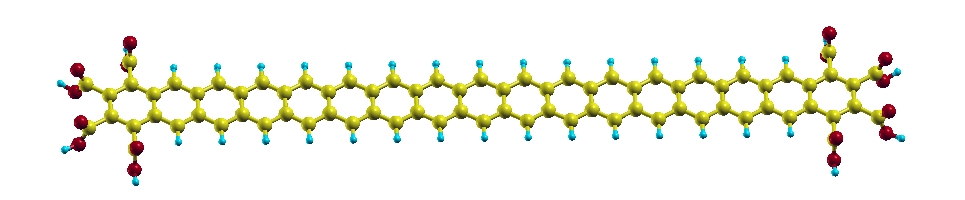}}

\end{tabular}
\caption{Atomic structures of molecules mentioned in Table 2 of the paper.}
\end{figure*}

\end{document}